\documentclass[aps,pra,preprint,groupedaddress,showpacs]{revtex4}

\usepackage{graphicx}
\usepackage{verbatim}
\usepackage{amsfonts} 
\usepackage{amsmath}
\usepackage{mathbbol}
\usepackage{amssymb}   
\usepackage{color}
\usepackage{bm}

\newcommand{\be}{\begin{equation}}
\newcommand{\ee}{\end{equation}}
 
 \definecolor{BrickRed}{cmyk}{0,0.89,0.94,0.28}
\definecolor{MidnightBlue}{cmyk}{0.98,0.13,0,0.43}
\definecolor{DarkGreen}{rgb}{0,0.7,0.1}

\begin{document}






\title{Casimir and Casimir-Polder Interactions for Magneto-dielectric Materials: Surface Scattering Expansion}



\author{ Giuseppe Bimonte$^{1,2}$, and Thorsten Emig$^{3}$}

\affiliation{${}^{1}$Dipartimento di Fisica E. Pancini, Universit\`{a} di
Napoli Federico II, Complesso Universitario
di Monte S. Angelo,  Via Cintia, I-80126 Napoli, Italy}
\affiliation{${}^{2}$INFN Sezione di Napoli, I-80126 Napoli, Italy}
\affiliation{${ }^{3}$ Laboratoire de Physique
Th\'eorique et Mod\`eles Statistiques, CNRS UMR 8626,
Universit\'e Paris-Saclay, 91405 Orsay cedex, France}

\email{giuseppe.bimonte@na.infn.it,
thorsten.emig@cnrs.fr}

\date{\today}

\begin{abstract}
We develop a general multiple scattering expansion (MSE) for computing Casimir forces between magneto-dielectric bodies and Casimir-Polder forces between polarizable particles and magneto-dielectric bodies. The approach is based on fluctuating electric and magnetic surface currents and charges. The surface integral equations for these surface fields can be formulated in terms of surface scattering operators (SSO). We show that there exists an entire family of such operators. One particular member of this family is only weakly divergent and allows for a MSE that appears to be convergent for general magneto-dielectric bodies. We proof a number of properties of this operator, and demonstrate explicitly convergence for sufficiently low and high frequencies, and for perfect conductors. General expressions are derived for the Casimir interaction between macroscopic bodies and for the Casimir-Polder interaction between particles and macroscopic bodies in terms of the SSO, both at zero and finite temperatures. An advantage of our approach above previous scattering methods is that it does not require the knowledge of the scattering amplitude (T-operator) of the bodies. A number of simple examples are provided to demonstrate the use of the method. Some applications of our approach have appeared previously [T.~Emig, G.~Bimonte, Phys.~Rev.~Lett.~{\bf 130}, 200401 (2023)]. Here we provide additional technical aspects and details of our approach.
\end{abstract}


\maketitle

\section{Introduction}

It is a quite common situation in physics, biology and chemistry to find surfaces of macroscopic objects and particles in close proximity to each other. Although these structures carry often no charge, they still experience a long-ranged interaction which results from modifications of the quantum and thermal fluctuations of the electromagnetic field by the objects. A well-known manifestation of this interaction is the Casimir force between two parallel perfectly conducting plates \cite{casimir}. Microscopically, this interaction can be understood as a collective, non-additive force between induced dipoles in the bodies. Indeed, the connection between an atomistic description and non-ideal macroscopic dielectric materials was established by Lifshitz who considered random currents within the interacting bodies to obtain the Casimir force between planar bodies \cite{lifshitz}.  This approach has been the core theory for interpreting most of the precision measurements of Casimir interactions between various materials and surface shapes which were enabled by an enormous progress in force sensing techniques and the fabrication of nano-structures \cite{lamoreaux,mohideen,chan,bressi,Decca:2003yb,Munday:2009xw,Sushkov:2011ik,Tang:2017kz,Bimonte:2016cr}.  Naturally, in practice macroscopic bodies have curved or structured surfaces. Hence, an approximation by planar surfaces is often not justified. 
Indeed, recent experiments \cite{Banishev:2013zp,Intravaia:2013yf,Wang2021} have demonstrated large deviations from common proximity approximations  \cite{Derjaguin:1934hb}, making theoretical formulations for a precise force computation highly desirable.  

An exact computation of Casimir forces in non-planar geometries is extremely hard. 
To date, the only non-planar configurations for which the force can be computed exactly are the  sphere-plate and the sphere-sphere systems, for Drude conductors in the high temperature limit \cite{bimonte2012ter,Ingold2021}. In principle, there exist methods to compute Casimir forces in arbitrary geometries. However, they are often limited in its practical applicability. Indeed, 
 enormous efforts have been put forward by many groups to develop theoretical and numerical methods that can cope with more general surface shapes \cite{Rodriguez:2011df,bimonte2017,bimonte2022}.  Specifically, the scattering method \cite{emig2007,kenneth2008,rahi2009}, originally devised for mirrors \cite{Genet03,lambrecht}, expresses the interaction between dielectric bodies in terms of their scattering amplitude, known as T-operator. While this approach has enabled most of recent theoretical progress, the T-operator is known only for highly symmetric bodies, such as sphere and cylinder, or for a few perfectly conducting shapes \cite{maghrebi}, practically exhausting this method. This scattering approach can be augmented by advanced numerical methods, for example for gratings \cite{Messina:2017fb}, but they can be limited by computational power required for convergence. A more fundamental limitation is that interlocked geometries evade this method 
due to lack of convergence of the partial wave expansions \cite{Wang2021}. If the surface is only gently curved, a gradient expansion can be used to obtain first order corrections to the proximity approximation \cite{fosco,bimonte2012bis}. The theoretical treatment of non-ideal materials with sharp surface features, such as used in atomic force microscopy or fabricated by lithographical techniques, is beyond the scope of existing methods. 

Substantial progress has been made over the last decade with fully numerical methods to compute Casimir forces for general shapes and materials. An important example is an approach based on a boundary element method (SCUFF-EM) for computing the interaction of fluctuating surface currents \cite{reid2013,rodriguez}. It is believed that this approach can provide in principle the exact force for arbitrary shapes, with computational power the only but practically important limiting factor \cite{Wang2021}. This method depends on a suitable refinement of the surface mesh for a broad band of relevant wave lengths. Therefore, the numerical effort for keeping discretization errors sufficiently small can be challenging. To the best of our knowledge, complementary, not fully numerical methods with comparably broad application range do not exist to date.

Here we develop a novel approach for computing Casimir forces for magneto-dielectric bodies of arbitrary shape. Conceptional, the Casimir force is related to fluctuating electric and magnetic surface currents and charges by the fluctuation-dissipation theorem \cite{Agarwal:1975sh}. This allows for a formulation of a general theory for Casimir forces that is based on scattering operators which are localized only on the surfaces of the interacting bodies. The important new features of our method are the following: 
(i) No knowledge of the scattering amplitude (T-operator) of the bodies is required. Hence, an important practical problem of the existing scattering approaches is overcome. (ii) No expansion of the EM field in partial waves, or expansion of currents in multipoles, is required. This eliminates the problems of convergence in geometries where surfaces interlock. (iii) Explicit expressions for the surface scattering operators are given in terms of free Green functions. (iv) Any basis for the tangential surface currents can be used, simplifying the computation of surface integrals appearing in the operator products. (v) The Casimir interaction can be expanded in the number of surface scatterings, leading to rapidly converging estimate for the interaction energy.

The general multiple scattering expansion is enabled by treating the back and forth scatterings of waves between different objects on an equal footing as the scatterings within an isolated object, eliminating the necessity to resort to the concept of a T-operator. In this formulation, a wave propagates freely in a magneto-dielectric medium between successive scattering points on the surfaces, no matter if the points belong to different objects or the same object. For perfectly conducting objects, in a seminal work Balian and Duplantier had demonstrated the very existence and convergence of a multiple scattering expansion for Casimir forces \cite{balian1977,balian1978}.
Our approach shows that a conceptional similar theory can be developed for 
for arbitrary dissipative magneto-dielectric materials. We provide a number of simple examples 
which show rapid convergence in the number of scatterings even at short surface separations. 
 Our work represents a powerful approach to substantially extend accurate predictions of Casimir forces to materials and shapes for which only computationally intensive fully numerical methods were available. 

A brief report of our findings  has appeared previously \cite{emig2023}. Here we provide  details of the derivation of the multiple scattering expansion and derive some important properties of the surface scattering operator (SSO). The paper is organized as follows. 
In Sec.~II we  derive the general expression of the SSO for a collection of $N$ magneto-dielectric bodies of any shape, placed at  arbitrary relative positions in space. 
In Sec.~III we express the Casimir interaction of two bodies, and the Casimir-Polder interaction between a polarizable particle and general magneto-dielectric body in terms of the SSO.
Several equivalent formulations of the SSO are discussed in Sec.~IV.
The limits of perfect conductors, and high and low frequencies are analyzed in Sec.~V.
In Sec.~VI we address the convergence properties of the MSE in general.
A number of simple examples demonstrate the application of the MSE in Sec.~VII.
 In Sec.~VIII we present our conclusions and a discussion of future applications of the MSE. Finally, several Appendices provide further technical details.

\section{Electric and magnetic surface currents from a multiple scattering expansion}

\begin{figure}
\includegraphics[width=0.6\textwidth]{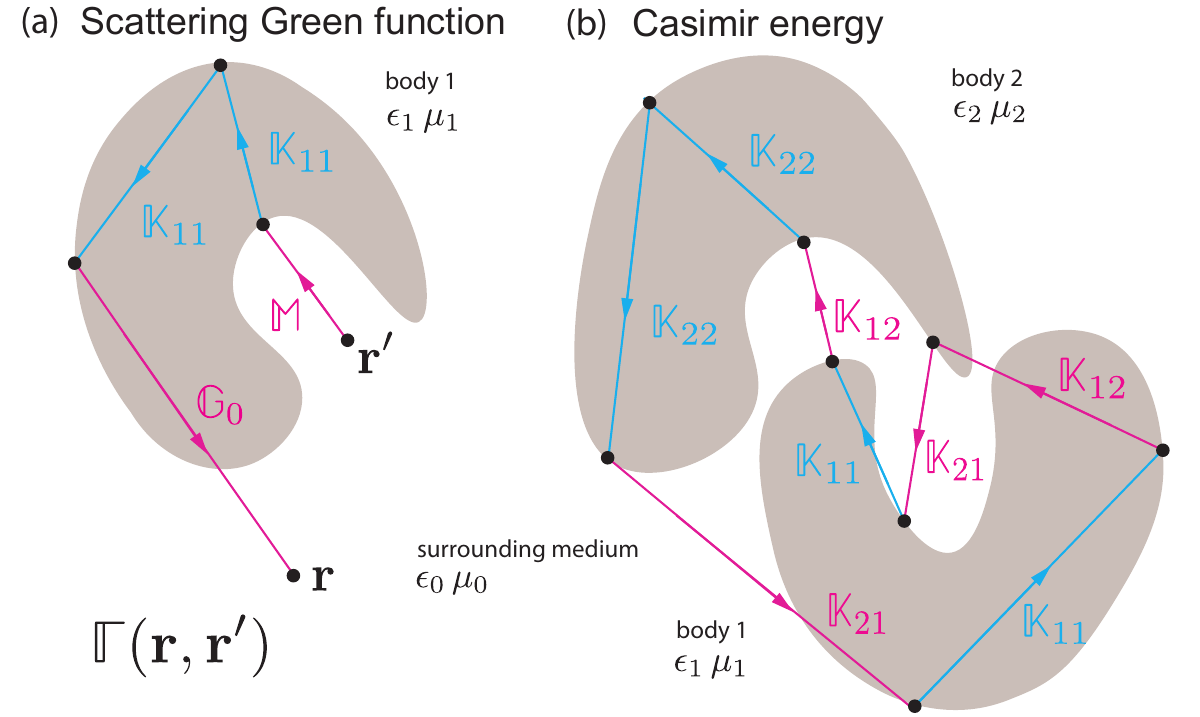} 
\caption{\label{fig:1}
{\bf Multiple scattering expansion.} Diagrammatic representation of contributions to the MSE, shown in panel {\bf a} for the scattering Green  function $\mathbb{\Gamma}({\bf r},{\bf r}')$ of a single body with source point ${\bf r}'$ and observation point ${\bf r}$, and in panel {\bf b} for the Casimir energy between two bodies. In the displayed examples, lines with arrows represent free propagation between surface points of the same body (blue lines) and to external points or between surface points of different bodies (magenta lines). Each free propagation between two surface points, followed by a scattering, is described by a surface operator $\mathbb{K}_{\sigma\sigma'}$. The bodies have dielectric and magnetic permittivities $\epsilon_1,\,\mu_1$ and $\epsilon_2,\,\mu_2$, respectively, and they are surrounded by a medium with permittivities $\epsilon_0,\,\mu_0$. $\mathbb{G}_0$ is the free Green tensor of the surrounding medium, and $\mathbb{M}$ described the tangential surface components of the incident field generated by a source at position ${\bf r}'$.}
\end{figure}

Before considering Casimir interactions, we first develop in this section the concept of surface currents and show how they naturally lead to an expansion of the electromagnetic (EM) field in the number of surface scatterings. This shall enable us to formulate a scattering expansion for the scattering Green tensor $\mathbb{\Gamma}({\bf r},{\bf r}')=\mathbb{G}({\bf r},{\bf r}')-\mathbb{G}_0({\bf r},{\bf r}')$, where  $\mathbb{G}$ is the $N$-body EM Green tensor and $\mathbb{G}_0$ is the empty space Green tensor for a homogenous medium with contrast $\epsilon_0$, $\mu_0$ (see App.~E). Physically, $\mathbb{\Gamma}({\bf r},{\bf r}')$ describes the {\it modification} of the EM field at position ${\bf r}$, due to the presence of the bodies, when it is generated by a source at position ${\bf r}'$. This naturally implies to construct $\mathbb{\Gamma}$ from the surface fields which are induced by an external source at the bodies. However, the primary current induced directly by the source induces in turn a secondary current, which induces again higher order currents, leading to an infinite sequence of induction processes. As we shall demonstrate subsequently, an exact mathematical description of these processes is provided by our multiple scattering expansions (MSE) for $\mathbb{\Gamma}$. While Green functions have been constructed in terms of surface currents, the existence and convergence of a MSE  between magneto-dielectric bodies is not obvious, particularly for Casimir interactions, and to the best of our knowledge had been demonstrated only for perfect electric conductors \cite{balian1977,balian1978,balian2004}. The MSE is based on surface integral equations that determine the tangential electric and magnetic fields at the surfaces $S_\sigma$ which can be considered as magnetic surface currents ${\bf m}_\sigma$  and electric surface currents ${\bf j}_\sigma$, acting as equivalent sources for the scattered field \cite{Harrington:2001yb}. This can be viewed as a mathematical reformulation of Huygens principle.
We note that in the static limit, it shall turn out that it is sufficient to consider the normal components of the EM field at the surfaces, corresponding to electric and magnetic surface charge densities. For finite frequencies, these charge densities are related to the surface currents by surface continuity equations.

In the following, we consider a configuration of $N$ material bodies with dielectric and magnetic permittivities $\epsilon_\sigma$ and $\mu_\sigma$ ($\sigma=1,\ldots,N$). The bodies are bounded by closed surfaces $S_\sigma$ which can be of arbitrary shape and separate their bulk from the surrounding homogeneous medium with dielectric and magnetic permittivities $\epsilon_0$ and $\mu_0$ \footnote{Surfaces which are open at infinity, such as an infinite plate or an infinite cylinder, are permitted as long as they separate the space into interior and exterior regions.}. 
From the uniqueness of an EM field in a region specified by sources within the region and the tangential components of the field over the boundary of the region, one can construct the total EM field $({\bf E}, {\bf H})$ separately in the region external to the bodies, and inside the $N$ interior regions of the bodies. When doing so, one can vary the field outside a given region at will as long as the surface currents are adjusted according to the jump conditions ${\bf j} = {\bf n} \times ({\bf H}_+ - {\bf H}_-)$, ${\bf m} = - {\bf n} \times ({\bf E}_+ - {\bf E}_-)$ where ${\bf n}$ is the surface normal pointing to the outside and the label $+(-)$ indicates the value when surface is approached from the outside (inside). To proceed, we make the choice that the field outside a given region vanishes as this allows us to replace the magneto-dielectric media outside the region by the medium inside the region, so that the surface currents on the boundary of the region radiate in homogenous unbounded space. Hence the field can be expressed in the interior of the bodies as the surface integral 
\begin{equation}
\label{eq:EM_sigma}
({\bf E}^{(\sigma)},{\bf H}^{(\sigma)})({\bf r})= \int_{S_\sigma} ds_{{\bf u}} \, \mathbb{G}_\sigma({\bf r},{\bf u}) ({\bf j}_{\sigma-},{\bf m}_{\sigma-})({\bf u})
\end{equation}
 where $\mathbb{G}_\sigma$ in the free Green tensor in a medium with permittivities $\epsilon_\sigma$, $\mu_\sigma$, and ${\bf j}_{\sigma-}=-{\bf n}_\sigma \times {\bf H}_-$, ${\bf m}_{\sigma-}={\bf n}_\sigma \times {\bf E}_-$ are the tangential fields when $S_\sigma$ is approached from the inside of the bodies. Exterior to the bodies the field 
\begin{equation}
\label{eq:EM_0}
 ({\bf E}^{(0)},{\bf H}^{(0)})({\bf r})= \int d{\bf r}' \, \mathbb{G}_0({\bf r},{\bf r}') ({\bf J},{\bf M})({\bf r}') + \sum_{\sigma=1}^N \int_{S_\sigma} ds_{{\bf u}} \, \mathbb{G}_0({\bf r},{\bf u}) ({\bf j}_{\sigma+},{\bf m}_{\sigma+})({\bf u})
\end{equation}
where now ${\bf j}_{\sigma+}={\bf n}_\sigma \times {\bf H}_+$, ${\bf m}_{\sigma+}=-{\bf n}_\sigma \times {\bf E}_+$ are the tangential fields when $S_\sigma$ is approached from the outside of the bodies
and we assumed an external source of electric and magnetic currents $({\bf J}, {\bf M})$ outside the bodies to generate the incident field $({\bf E}_{\rm inc},{\bf H}_{\rm inc})$.
Surface integral equations for the surface fields follow by taking advantage of the property of the surface integrals that they are also defined when ${\bf r}$ is located {\it on} the surfaces and their corresponding value is the average of the limits taken from the inside and the outside \cite{muller}, and that one of the two limits vanishes by construction, leading to 
\begin{align}
\begin{split}
({\bf m}_{\sigma-},-{\bf j}_{\sigma-}) ({\bf u}) & = 2 {\bf n}_\sigma({\bf u}) \times ({\bf E}^{(\sigma)},{\bf H}^{(\sigma)})({\bf u}) \\
({\bf m}_{\sigma+},-{\bf j}_{\sigma+})({\bf u}) & = - 2 {\bf n}_\sigma({\bf u}) \times ({\bf E}^{(0)},{\bf H}^{(0)})({\bf u})
\end{split}
\end{align}
for ${\bf u}$ located on surface $S_\sigma$. Associated with the surface currents must be surface charges which we are given by the (rescaled) surface charge densities, defined on both sides of the surfaces as
\begin{align}
\begin{split}
(\varrho_{j,\sigma-},\varrho_{m,\sigma-}) ({\bf u}) & = -2 {\bf n}_\sigma({\bf u})  ({\bf E}^{(\sigma)},{\bf H}^{(\sigma)})({\bf u}) \\
(\varrho_{j,\sigma+},\varrho_{m,\sigma+})({\bf u}) & =  2 {\bf n}_\sigma({\bf u})  ({\bf E}^{(0)},{\bf H}^{(0)})({\bf u}) \, .\label{surchar}
\end{split}
\end{align}
Finally, to couple the interior and exterior solutions, we impose the usual continuity conditions on the tangential components of $({\bf E} ,{\bf H})$ at the interfaces between different media, leading to one unique set of surface currents $({\bf j}_\sigma,{\bf m}_\sigma) \equiv ({\bf j}_{\sigma+},{\bf m}_{\sigma+}) =- ({\bf j}_{\sigma-},{\bf m}_{\sigma-})$. Similarly, imposing continuity on the normal components of ${\bf D}=\epsilon {\bf E}$ and ${\bf B} = \mu {\bf H}$ leads to the relation 
\begin{equation}
\label{eq:varrho_in_out_rel}
\varrho_{j,\sigma-} = -\frac{\epsilon_0}{\epsilon_\sigma} \varrho_{j,\sigma+} , \quad 
\varrho_{m,\sigma-} = -\frac{\mu_0}{\mu_\sigma} \varrho_{m,\sigma+}
\end{equation}
between the interior and exterior charge densities. Hence, it is sufficient to consider the unique set of surface charge densities $(\varrho_{j,\sigma},\varrho_{m,\sigma})\equiv(\varrho_{j,\sigma+},\varrho_{m,\sigma+})$. Since the field $({\bf E}^{(\sigma)},{\bf H}^{(\sigma)})$ obeys the source free Maxwell equations in the interior region of the surface $S_\sigma$, the interior surface currents and charges are related by the continuity equations
\begin{align}
\begin{split}
\nabla {\bf j}_{\sigma-} &= -\kappa \epsilon_\sigma \varrho_{j,\sigma-} \\
\nabla {\bf m}_{\sigma-} &= -\kappa \mu_\sigma \varrho_{m,\sigma-} \, ,
\end{split}
\end{align}
or, due to Eq.~(\ref{eq:varrho_in_out_rel}), equivalently by the continuity equations for the unique surface currents and charges
\begin{align}
\begin{split}
\label{eq:continuity_surface_charges}
\nabla {\bf j}_{\sigma} &= -\kappa \epsilon_0 \varrho_{j,\sigma} \\
\nabla {\bf m}_{\sigma} &= -\kappa \mu_0 \varrho_{m,\sigma} \, .
\end{split}
\end{align}
Now we have expressed the surface currents $({\bf j}_\sigma,{\bf m}_\sigma)$ and charges $(\varrho_{j,\sigma},\varrho_{m,\sigma})$ in terms of both the interior field $({\bf E}^{(\sigma)},{\bf H}^{(\sigma)})$ and the exterior field $({\bf E}^{(0)},{\bf H}^{(0)})$. This yields the surface integral equations
\begin{align}
\label{eq:IE1}
({\bf m}_{\sigma},-{\bf j}_{\sigma}) ({\bf u}) & = -2 {\bf n}_\sigma({\bf u}) \times ({\bf E}^{(\sigma)},{\bf H}^{(\sigma)})({\bf u}) \\
\label{eq:IE2}
({\bf m}_{\sigma},-{\bf j}_{\sigma})({\bf u}) & = - 2 {\bf n}_\sigma({\bf u}) \times ({\bf E}^{(0)},{\bf H}^{(0)})({\bf u})\\
\label{eq:IE3}
(\varrho_{j,\sigma},\varrho_{m,\sigma}) ({\bf u}) & = 2 {\bf n}_\sigma({\bf u})  \left( \frac{\epsilon_\sigma}{\epsilon_0}{\bf E}^{(\sigma)},\frac{\mu_\sigma}{\mu_0}{\bf H}^{(\sigma)}\right)({\bf u}) \\
\label{eq:IE4}
(\varrho_{j,\sigma},\varrho_{m,\sigma})({\bf u}) & =  2 {\bf n}_\sigma({\bf u})  ({\bf E}^{(0)},{\bf H}^{(0)})({\bf u}) \, .
\end{align}
where the fields are given by the integrals in Eqs.~(\ref{eq:EM_sigma}), (\ref{eq:EM_0}) with $({\bf j}_{\sigma+},{\bf m}_{\sigma+})=({\bf j}_\sigma,{\bf m}_\sigma)$ and $({\bf j}_{\sigma-},{\bf m}_{\sigma-})=-({\bf j}_{\sigma},{\bf m}_{\sigma})$.
These $8N$ surface integral equations constitute an overdetermined system for the $2N$ surface currents or tangential surface fields, and the $2N$ surface charge densities, which must be related to the surface currents by the continuity equations (\ref{eq:continuity_surface_charges}). Existence of a unique solution requires that only $4N$ equations are independent, agreeing with the number of constraints imposed by the continuity of the tangential and normal field components. The additional $4N$ constraints, implicitly fulfilled by construction of the fields, must account for the unique relation between the components of the electric and magnetic fields on both sides of the surfaces as specification of either tangential ${\bf E}$ or tangential ${\bf H}$ determine a unique solution to the exterior and interior problems. For this reason, a consistent set of $4N$ integral equations with a unique solution can be obtained by 
taking linear combinations of the set of $4N$ equations involving $({\bf E}^{(\sigma)},{\bf H}^{(\sigma)})$ {\it and} the corresponding set involving $({\bf E}^{(0)},{\bf H}^{(0)})$ but not by considering only one of the two sets as this would ignore the coupling of the interior and exterior fields.

We first consider the integral equations for the surface currents, Eqs.~(\ref{eq:IE1}), (\ref{eq:IE2}).  In general, when taking linear combinations of the integral equations, one can choose $4N$ suitable coefficients which form $2N$ diagonal $2\times 2$ matrices $\mathbb{C}^{i}_\sigma$, $\mathbb{C}^{e}_\sigma$ acting on the two field components of the interior and exterior integral equations. To interpret the integral equations as successive scatterings, we introduce the surface scattering operators (SSOs) $\mathbb{K}_{\sigma\sigma'}({\bf u},{\bf u}')$ which describe free propagation from ${\bf u}'$ on surface $S_{\sigma'}$ to ${\bf u}$ on surface $S_{\sigma}$ and scattering at point ${\bf u}$ 
\begin{equation}
\label{eq:1}
\mathbb{K}_{\sigma\sigma'}({\bf u},{\bf u}') = 2 \mathbb{P} (\mathbb{C}^{i}_\sigma+\mathbb{C}^{e}_\sigma)^{-1} {\bf n}_\sigma({\bf u}) \times  \left[  \delta_{\sigma\sigma'} \mathbb{C}^{i}_\sigma  \mathbb{G}_\sigma({\bf u},{\bf u}')- \mathbb{C}^{e}_\sigma \mathbb{G}_0({\bf u},{\bf u}')\right] \, , \quad \mathbb{P} = \big(\begin{smallmatrix} 0 & -1 \\ 1 & 0 \end{smallmatrix} \big)
\end{equation}
acting on electric and magnetic tangential surface fields at ${\bf u}'$ ($\delta_{\sigma\sigma'}$ is the Kronecker delta). With these SSOs the surface currents are determined in terms of the external source $({\bf J},{\bf M})$ by the Fredholm integral equations of the 2nd kind
\begin{equation}
\label{eq:Fredholm_currents}
\sum_{\sigma'=1}^N\int_{S_{\sigma'}}\!\!ds_{{\bf u}'} \,\left[\mathbb{1} - \mathbb{K}_{\sigma\sigma'}({\bf u},{\bf u}')\right] \big(\begin{smallmatrix} {\bf j}_{\sigma'} \\ {\bf m}_{\sigma'} \end{smallmatrix} \big) ({\bf u}')
= \!\int d{\bf r} \,\mathbb{M}_{\sigma}({\bf u},{\bf r})\big(\begin{smallmatrix} {\bf J} \\ {\bf M}\end{smallmatrix}\big) ({\bf r})
\end{equation}
with
\begin{equation}
\label{eq:3}
\mathbb{M}_\sigma({\bf u},{\bf r}) =  -2 \mathbb{P} (\mathbb{C}^{i}_\sigma+\mathbb{C}^{e}_\sigma)^{-1} \mathbb{C}^{e}_\sigma\, {\bf n}_\sigma({\bf u}) \times  \mathbb{G}_0({\bf u},{\bf r}) \, .
\end{equation}
(For  an alternative derivation of the SSO  we refer to App.~A.)
More explicit expressions for the SSO for different choices of the coefficient matrices will be given below in Sec.~IV.
As we shall see, the choice of coefficients $\mathbb{C}^{i}_\sigma$, $\mathbb{C}^{e}_\sigma$ provides a powerful tool to engineer convergence of the MSE. 
Uniqueness of the solution of the integral equation (\ref{eq:Fredholm_currents}) is ensured if  one can show that the operator ${\mathbb K}$ does not have an eigenvalue equal to one.  Such a proof for any (complex) frequency can be found in  the book \cite{muller}  for a particular choice of coefficients, denoted by (C1) in Sec.~IV below, and for a single body.  A simple generalization of the proof allows to show that the result remains true for any  number of bodies. After an appropriate re-scaling of the EM field,  one can  show  that the result holds also  for all values of the coefficients as long as ${\mathbb C}^{e}_{\sigma}+{\mathbb C}^{i}_{\sigma}$ is different from zero. 
Explicit computation of the SSO requires integration of the {\it free} space Green tensor in homogenous media over the bodies' surfaces which can be performed analytically in some cases. Contributions to the Casimir energy from scatterings between remote surface positions are exponentially damped with distance as we need to consider the Green tensor only for purely imaginary frequencies. 

Next, we consider the integral equations which determine the surface charge densities. 
While the electromagnetic scattering problem is basically solved in terms of surface currents determined by Eq.~(\ref{eq:Fredholm_currents}), it turns out that the zero frequency limit $\kappa=0$ requires a separate treatment due to a divergent term in the SSO for $\kappa\to 0$. The corresponding static problem is described in terms of surface charges only, as we shall see now.
We take linear combinations of the integral equations for the surface currents, Eqs.~(\ref{eq:IE3}), (\ref{eq:IE4}), with scalar interior coefficients $\mathbb{c}^{i}_{j,\sigma}$, $\mathbb{c}^{i}_{m,\sigma}$and exterior coefficients $\mathbb{c}^{e}_{j,\sigma}$, $\mathbb{c}^{e}_{m,\sigma}$. Using the surface divergence theorem and the continuity Eqs.~(\ref{eq:continuity_surface_charges}), one gets two Fredholm integral equations of the 2nd kind,
\begin{align}
\varrho_{j,\sigma}({\bf u}) & + \frac{2}{\mathbb{c}^{e}_{j,\sigma}+ \mathbb{c}^{i}_{j,\sigma}} \sum_{\sigma'=1}^N\int_{S_{\sigma'}}\!\!ds_{{\bf u}'} \,\left[
\kappa {\bf n}_\sigma({\bf u}) {\bf j}_{\sigma'}({\bf u}')\left( \mathbb{c}^{e}_{j,\sigma} \mu_0 g_0({\bf u}-{\bf u}') -\frac{\epsilon_\sigma}{\epsilon_0} \mathbb{c}^{i}_{j,\sigma} \mu_\sigma g_\sigma({\bf u}-{\bf u}') \delta_{\sigma\sigma'} \right) \right. \nonumber \\
& \left. + \varrho_{j,\sigma'}({\bf u}) \left(  \mathbb{c}^{e}_{j,\sigma} \partial_{{\bf n}_\sigma({\bf u})} g_0({\bf u}-{\bf u}')  - \mathbb{c}^{i}_{j,\sigma} \partial_{{\bf n}_\sigma({\bf u})} g_\sigma({\bf u}-{\bf u}') \delta_{\sigma\sigma'}\right) \right. \nonumber\\
& \left. + {\bf n}_\sigma({\bf u}) \left( \left( \mathbb{c}^{e}_{j,\sigma} \nabla_{\bf u} g_0({\bf u}-{\bf u}') - \frac{\epsilon_\sigma}{\epsilon_0} \mathbb{c}^{i}_{j,\sigma} \nabla_{\bf u} g_\sigma({\bf u}-{\bf u}') \delta_{\sigma\sigma'} \right) \times {\bf m}_{\sigma'}({\bf u}')\right)
\right] \nonumber\\
& = \frac{2 \mathbb{c}^{e}_{j,\sigma} }{\mathbb{c}^{e}_{j,\sigma} +\mathbb{c}^{i}_{j,\sigma} }  {\bf n}_\sigma({\bf u})  {\bf E}_{\rm inc}({\bf u})
\label{eq:IE_rho_j}
\\
\varrho_{m,\sigma}({\bf u}) & + \frac{2}{\mathbb{c}^{e}_{m,\sigma}+ \mathbb{c}^{i}_{m,\sigma}} \sum_{\sigma'=1}^N\int_{S_{\sigma'}}\!\!ds_{{\bf u}'} \,\left[
\kappa {\bf n}_\sigma({\bf u}) {\bf m}_{\sigma'}({\bf u}')\left( \mathbb{c}^{e}_{m,\sigma} \epsilon_0 g_0({\bf u}-{\bf u}') -\frac{\mu_\sigma}{\mu_0} \mathbb{c}^{i}_{m,\sigma} \epsilon_\sigma g_\sigma({\bf u}-{\bf u}') \delta_{\sigma\sigma'} \right) \right. \nonumber \\
& \left. + \varrho_{m,\sigma'}({\bf u}) \left(  \mathbb{c}^{e}_{m,\sigma} \partial_{{\bf n}_\sigma({\bf u})} g_0({\bf u}-{\bf u}')  - \mathbb{c}^{i}_{m,\sigma} \partial_{{\bf n}_\sigma({\bf u})} g_\sigma({\bf u}-{\bf u}') \delta_{\sigma\sigma'}\right) \right. \nonumber\\
& \left. - {\bf n}_\sigma({\bf u}) \left( \left( \mathbb{c}^{e}_{m,\sigma} \nabla_{\bf u} g_0({\bf u}-{\bf u}') - \frac{\mu_\sigma}{\mu_0} \mathbb{c}^{i}_{m,\sigma} \nabla_{\bf u} g_\sigma({\bf u}-{\bf u}') \delta_{\sigma\sigma'} \right) \times {\bf j}_{\sigma'}({\bf u}')\right)
\right] \nonumber\\
& = \frac{2 \mathbb{c}^{e}_{m,\sigma} }{\mathbb{c}^{e}_{m,\sigma} +\mathbb{c}^{i}_{m,\sigma} }  {\bf n}_\sigma({\bf u})  {\bf H}_{\rm inc}({\bf u})
\, ,
\label{eq:IE_rho_m}
\end{align}
where $g_\sigma$ is the scalar free Green function (see App.~E).   Different choices for the coefficients $\mathbb{c}^{i/e}_{j/m,\sigma}$ will be discussed in Sec.~IV.
Remarkably, there exists a choice of  the coefficients (see Eq.(\ref{eq:coeff_c1_static})) such that  in the static limit, $\kappa\to 0$, the above integral equations can be expressed in terms of the surface charges only, as we shall show in Sec.~V.A below. 

\section{Interactions due to fluctuations of the electromagnetic field / surface currents}

\subsection{Scattering Green tensor}

The scattering Green tensor $\mathbb{\Gamma}({\bf r},{\bf r}')=\mathbb{G}({\bf r},{\bf r}')-\mathbb{G}_0({\bf r},{\bf r}')$ is essential to compute the expectation value of the stress tensor, and hence Casimir forces. It is determined by the field generated by the surface currents, and hence 
\begin{equation}
\label{eq:Gamma}
\mathbb{\Gamma}({\bf r},{\bf r}')=
\int_S ds_{\bf u} \int_S ds_{{\bf u}'}\,  \mathbb{G}_0({\bf r},{\bf u}) (\mathbb{1}-\mathbb{K})^{-1}({\bf u},{\bf u}') \mathbb{M}({\bf u}',{\bf r}')
\end{equation}
where the integration extends over all surfaces $S_\sigma$ and a summation over all  surface labels $\sigma$ is understood. The operator $\mathbb{M}({\bf u}',{\bf r}')$ is proportional to the free Green tensor,
\begin{equation}
\label{eq:M_decomp}
\mathbb{M}({\bf u}',{\bf r}') = \mathbb{V}({\bf u}') \mathbb{G}_0({\bf u}',{\bf r}')\, ,\quad 
 \mathbb{V}({\bf u}') = -2 \mathbb{P} (\mathbb{C}^{i}_\sigma+\mathbb{C}^{e}_\sigma)^{-1} \mathbb{C}^{e}_\sigma\, {\bf n}_\sigma({\bf u}') \times \cdot \, ,
\end{equation}
where $\cdot$ is a placeholder for the argument on which the operator acts.
The existence of a MSE follows from the Fredholm type of the operator $(\mathbb{1}-\mathbb{K})^{-1}$ that permits an expansion in powers of $\mathbb{K}$ \footnote{The series expansion of $(\mathbb{1}-\mathbb{K})^{-1}$  powers of $\mathbb{K}$ is known in the mathematical literature as the Neumann series} and hence in the number of scatterings, as illustrated for one body in Fig.~\ref{fig:1}(a). 

\subsection{Casimir force between magneto-dielectric bodies}

We now derive the Casimir interaction among the bodies. Following the method in \cite{bimonte2021}, we first express the Casimir force ${\bf F}_\sigma$ on one of the bodies, labelled by $\sigma$, as the integral of the expectation value of the EM stress tensor at discrete Matsubara imaginary frequencies $\xi=i\omega$ with $\xi=\xi_n=2\pi n k_B T/\hbar$ with $n=0,1,\ldots$,  over the surface $S_\sigma$ using the fluctuation-dissipation theorem. A divergence in the surface integral, originating from the empty space stress tensor and hence unrelated to the Casimir force, is readily removed by replacing the $N$-body EM Green tensor $\mathbb{G}$ by the scattering Green tensor $\mathbb{\Gamma}({\bf r},{\bf r}')$.

The regularized stress tensor involves only $\mathbb{\Gamma}$, and it can be shown  \cite{bimonte2021} that the Casimir force on body $\sigma$ is determined by the operator $(\mathbb{1}-\mathbb{K})^{-1}({\bf u},{\bf u}')\mathbb{V}({\bf u}')$ which is sandwiched between the free Green tensors in the scattering Green tensor, see Eq.~(\ref{eq:Gamma}). Hence, the Casimir force is given by
\begin{equation}
\label{eq:Force_1}
{\bf F}_\sigma =  k_B T \sideset{}{'}\sum_{n=0}^\infty {\rm Tr} [(\mathbb{1}-\mathbb{K})^{-1} \mathbb{V} \, \nabla_{{\bf r}_\sigma} {\mathbb G}_0] \, .
\end{equation}
Due to the important general relation
\begin{equation}
\label{eq:rel_nabla_K}
\nabla_{{\bf r}_\sigma} {\mathbb K} = \mathbb{V} \, \nabla_{{\bf r}_\sigma} {\mathbb G}_0
\end{equation}
the force can be written solely in terms of the SSO, expressed as a sum over Matsubara frequencies 
$\xi_n$ by  
\begin{equation}
\label{eq:Force_2}
{\bf F}_\sigma =  k_B T \sideset{}{'}\sum_{n=0}^\infty {\rm Tr} [ (\mathbb{1}-\mathbb{K})^{-1} \nabla_{{\bf r}_\sigma} {\mathbb K}] 
\end{equation}
where $\nabla_{{\bf r}_\sigma}$
is the gradient with respect to the position of the body, and the bare Casimir energy assumes the simple expression 
 \be 
 {\cal E} =  k_B T \sideset{}{'}\sum_{n=0}^\infty {\rm Tr} \log (\mathbb{1}-\mathbb{K})\;,
 \ee 
 (where the primed sum gives a weight of $1/2$ to the $n=0$ term). Here the trace Tr involves a sum over vector indices of the electric and magnetic components and an integration over all surfaces.
To gain insight into the structure of the MSE for the Casimir energy, we consider two bodies. After subtracting the self-energies, arising from isolated scatterings on a single body, the energy is expressed in terms of four SSO as
\begin{equation}
\label{eq:4}
{\cal E} = k_B T \sideset{}{'}\sum_{n=0}^\infty {\rm Tr} \log \left[ \mathbb{1} - (\mathbb{1}-\mathbb{K}_{11})^{-1} \mathbb{K}_{12}(\mathbb{1}-\mathbb{K}_{22})^{-1}\mathbb{K}_{21}\right] \, .
\end{equation}
We note that this formula provides the exact representation of the Casimir energy for all allowed choices of the coefficients $\mathbb{C}^{i}_\sigma$, $\mathbb{C}^{e}_\sigma$ (see also next section). 
After expanding both the logarithm and the inverse operators in powers of the SSOs we obtain the MSE which involves at least one scattering on each body with closed paths going from body 1 to body 2 and back ($\mathbb{K}_{12}$ and $\mathbb{K}_{21}$), possibly multiple times, and with an arbitrary number (including zero) of scatterings on each body ($\mathbb{K}_{11}$ and $\mathbb{K}_{22}$), as illustrated in Fig.~\ref{fig:1}(b). Comparison with scattering approaches relying on the knowledge of the bodies T-matrix shows that our MSE constructs the T-matrix in the number of scatterings on individual bodies by expanding $(\mathbb{1}-\mathbb{K}_{\sigma\sigma})^{-1}$, treating scatterings inside individual bodies and between them on an equal footing.  It is important to compare the MSE  with the so-called Born series expansion of the Green's tensor  \cite{buhmann2, Buhmann:2006aa}, which is an expansion in terms of iterated integrals over the {\it volumes} occupied by the bodies. Since  our MSE is instead an expansion in terms of iterated integrals over the bodies {\it surfaces},   it is clear that compared with the Born expansion, the MSE saves an enormous amount  of computing time, especially when high orders are considered.   We note also that while the Born series is an expansion in the dielectric contrast, our MSE is instead an expansion in the number of scatterings. 

Previously, scatterings of EM waves at dielectric media have been described in terms of electric and magnetic surface currents for real frequencies, revealing sometimes poor convergence of expansions in the number of scatterings. However,  since Casimir interactions can be formulated in terms of correlations of the EM field for purely imaginary frequencies, the exponential decay of Green tensors in separation can be expected to lead to rather fast convergence of the MSE for the scattering the Green function and the Casimir energy. 
This had been demonstrated only for perfect electric conductors, bases on a MSE that ignores the coupling between electric and magnetic surface currents \cite{balian1977}. One remarkable property of this previous approach, the cancellation of an odd overall number of scatterings, is explained in retrospect by our general MSE by the observation that ignorance of the coupling leads to SSO with opposite signs for the electric and magnetic components. 

\subsection{Casimir-Polder force between a polarizable particle and a magneto-dielectric body}

The Casimir-Polder interaction between a polarizable particle and an magneto-dielectric body can be obtained as a simple byproduct of our general approach. We assume that the particle is characterized by a frequency dependent electric polarizability tensor ${\bm \alpha}(\omega)$ and a magnetic polarizability tensor ${\bm \beta}(\omega)$. The classical energy of an induced dipole is then given by
\begin{equation}
\label{eq:inducedDipole}
{\cal E}_\text{cl} = -\frac{1}{2} \sum_{i,j=1}^{3} \left[ \alpha_{ij} E_i E_j + \beta_{ij} H_i H_j\right] \, .
\end{equation}
Using the fluctuation-dissipation theorem, this expression is averaged over EM field fluctuations. After removing a divergent contribution from empty space, the Casimir Polder energy is expressed in terms of the scattering Green tensor as
\begin{equation}
\label{eq:CPenergy}
{\cal E}_\text{CP} = - 4\pi k_B T \, \sideset{}{'}\sum_{n=0}^\infty  \kappa_n \, \sum_{i,j=1}^{3} \left[ \alpha_{ij} ({\rm i} \, \xi_n) \mathbb{\Gamma}^{(EE)}_{ij}({\bf r}_0,{\bf r}_0;\kappa_n)+ \beta_{ij} ({\rm i} \, \xi_n) \mathbb{\Gamma}^{(HH)}_{ij}({\bf r}_0,{\bf r}_0;\kappa_n) \right] \, ,
\end{equation}
where we assumed that the particle is located at position ${\bf r}_0$. 
Substitution of $\mathbb{\Gamma}$ from Eq.~(\ref{eq:Gamma}) yields the interaction energy of the particle with a body in terms of the SSO. This energy can be computed by a MSE with respect to the number of scatterings at the surface of the body. It is instructive to write down explicitly the first terms of the scattering expansion of the Casimir-Polder energy, assuming for simplicity that the electric polarizability of the particle is isotropic $\alpha_{ij}=\alpha \,\delta_{ij}$ , and that its magnetic polarizability $\beta$ is negligible:
\begin{align}
{\cal E}_\text{CP} &= - 4\pi k_B T \, \sideset{}{'}\sum_{n=0}^\infty  \kappa_n \,  \alpha({\rm i} \, \xi_n)\, \left\{ \sum_{p={E,H}} \int_S ds_{\bf u}  {\rm tr} \left[\mathbb{G}_{0}^{(E p)}({\bf r}_0,{\bf u}; \kappa_n)    \mathbb{M}^{(p E)}({\bf u},{\bf r}_0; \kappa_n) \right] \right. \nonumber \\
&\left.+\sum_{p,q={E,H}} \int_S ds_{\bf u}  \int_S ds_{{\bf u}'}  {\rm tr} \left[\mathbb{G}_{0}^{(E p)}({\bf r}_0,{\bf u}; \kappa_n)   \mathbb{K}^{(p q)}({\bf u},{\bf u}'; \kappa_n)   \mathbb{M}^{(q E)}({\bf u}',{\bf r}_0; \kappa_n) \right] \right\}+\cdots \label{CPser}
\end{align}
where ${\rm tr}$ denotes a trace over tensor spatial indices. Recalling that the kernels $\mathbb{K}({\bf u},{\bf u}')$ and $\mathbb{M}({\bf u},{\bf r})$ are combinations of free-space Green tensors $\mathbb{G}_{0}$ and $\mathbb{G}_{\sigma}$, and that the latter are elementary functions, we see from the above equation that the CP energy is expressed in terms of iterated integrals of elementary functions extended on the surface $S$ of the body. Since for imaginary frequencies the Green tensors decay exponentially with distance, Eq. (\ref{CPser}) makes evident the intuitive fact that the points of the surface that  are closest to the particle dominate the interaction.  

\section{Equivalent formulations of the SSO}
\label{sec:equiform}

With different interior coefficient matrices $\mathbb{C}^{i}_\sigma$ and exterior coefficient matrices $\mathbb{C}^{e}_\sigma$ the SSO form an equivalence class of operators in the sense that Eq.~(\ref{eq:Fredholm_currents}) yields the same surface currents for a given external source for all coefficients, as long as neither the interior nor the exterior matrices vanish for any $\sigma$, and the sum $\mathbb{C}^{i}_\sigma+\mathbb{C}^{e}_\sigma$ is invertible. Consequently, the scattering Green tensor and the Casimir energy must be also independent of the choice made for the coefficients. The surface currents and the Casimir energy at any {\it finite} order of the MSE, however, in general do depend on the chosen coefficients, and hence does the rate of convergence of the MSE. This remarkable property provides an effective method to optimize convergence for different permittivities and even frequencies by suitable adjustment of coefficients. 

Physically, the required relation between the tangential surface fields ${\bf n}_\sigma \times {\bf E}$ and  ${\bf n}_\sigma \times {\bf H}$ is in general obeyed only approximately at any finite order of the MSE, with the approximation converging to the exact relation with  increasing MSE order. Indeed, at first order, ${\bf E}$  and ${\bf H}$ of the incident field are rescaled differently at each body by the chosen coefficients $\mathbb{C}^{i}_\sigma$, $\mathbb{C}^{e}_\sigma$, see Eq.~(\ref{eq:3}). The coefficients hence set the initial field for the MSE iteration and they control how the exact tangential surface fields are build up successively by the MSE. 

Among the infinitely many choices there are a few which we consider important to discuss explicitly and for which we shall provide detailed expressions of the SSO's.

\noindent
{\bf (C1)}  In general, the SSO has a leading singularity that diverges as $1/|{\bf u}-{\bf u}'|^\gamma$ with $\gamma=3$ when the two surface positions ${\bf u}$, ${\bf u}'$ approach each other. 
There exists a choice of coefficients \cite{muller}, however, for which the singularity is reduced to a weaker divergence with exponent $\gamma=1$, presumably accelerating convergence. 
The coefficient matrices are 
\begin{equation}
\mathbb{C}^{i}_\sigma={\rm diag}(\epsilon_\sigma,\mu_\sigma)\, , \quad
\mathbb{C}^{e}_\sigma={\rm diag}(\epsilon_0,\mu_0) \, .
\end{equation}
The corresponding explicit expressions of the SSO's $\mathbb{K}$ and of the operator $\mathbb{M}$  read
\begin{equation}
\begin{aligned}
\label{eq:12}
\mathbb{K}^{(EE)}_{\sigma\sigma'}({\bf u},{\bf u}') & = \frac{2}{\mu_0+\mu_\sigma} {\bf n}_\sigma({\bf u}) \times \left[ \mu_0 \mathbb{G}_0^{(HE)}({\bf u},{\bf u}') - \delta_{\sigma\sigma'} \mu_\sigma \mathbb{G}_\sigma^{(HE)}({\bf u},{\bf u}')\right] \\
\mathbb{K}^{(HH)}_{\sigma\sigma'}({\bf u},{\bf u}') & = \frac{2}{\epsilon_0+\epsilon_\sigma} {\bf n}_\sigma({\bf u}) \times \left[ -\epsilon_0 \mathbb{G}_0^{(EH)}({\bf u},{\bf u}') + \delta_{\sigma\sigma'} \epsilon_\sigma \mathbb{G}_\sigma^{(EH)}({\bf u},{\bf u}')\right] \\
\mathbb{K}^{(EH)}_{\sigma\sigma'}({\bf u},{\bf u}') & = \frac{2}{\mu_0+\mu_\sigma} {\bf n}_\sigma({\bf u}) \times \left[ \mu_0 \mathbb{G}_0^{(HH)}({\bf u},{\bf u}') - \delta_{\sigma\sigma'} \mu_\sigma \mathbb{G}_\sigma^{(HH)}({\bf u},{\bf u}')\right] \\
\mathbb{K}^{(HE)}_{\sigma\sigma'}({\bf u},{\bf u}') & = \frac{2}{\epsilon_0+\epsilon_\sigma} {\bf n}_\sigma({\bf u}) \times \left[ - \epsilon_0 \mathbb{G}_0^{(EE)}({\bf u},{\bf u}') + \delta_{\sigma\sigma'} \epsilon_\sigma \mathbb{G}_\sigma^{(EE)}({\bf u},{\bf u}')\right] 
\end{aligned} \;,
\end{equation}
and
\begin{equation}
\begin{aligned}
\label{eq:13}
\mathbb{M}^{(EE)}_{\sigma}({\bf u},{\bf r}) & = \frac{2\mu_0}{\mu_0+\mu_\sigma} {\bf n}_\sigma({\bf u}) \times \mathbb{G}_0^{(HE)}({\bf u},{\bf r})\,, \\
\mathbb{M}^{(EH)}_{\sigma}({\bf u},{\bf r})  & = \frac{2\mu_0}{\mu_0+\mu_\sigma} {\bf n}_\sigma({\bf u}) \times \mathbb{G}_0^{(HH)}({\bf u},{\bf r})\\
\mathbb{M}^{(HE)}_{\sigma}({\bf u},{\bf r}) & = - \frac{2\epsilon_0}{\epsilon_0+\epsilon_\sigma} {\bf n}_\sigma({\bf u}) \times \mathbb{G}_0^{(EE)}({\bf u},{\bf r})\,, \\
\mathbb{M}^{(HH)}_{\sigma}({\bf u},{\bf r})  &= - \frac{2\epsilon_0}{\epsilon_0+\epsilon_\sigma} {\bf n}_\sigma({\bf u}) \times \mathbb{G}_0^{(EH)}({\bf u},{\bf r}) \, ,
\end{aligned}
\end{equation}
with the free Green tensor $\mathbb{G}_\sigma$ which can be found in App.~E. Substitution of this tensor yields the more explicit form in terms of the scalar Green functions $g_\sigma({\bf u}-{\bf u}')$,
\begin{equation}
\begin{aligned}
\label{eq:K_explicit}
{\mathbb K}^{(EE)}_{\sigma\sigma'}({\bf u},{\bf u}') &= \frac{2}{\mu_0+\mu_\sigma} \Big[ {\bf n}({\bf u})  \times \big( \cdot \times \nabla ( - \mu_0 g_0({\bf u}-{\bf u}') + \delta_{\sigma\sigma'} \mu_\sigma  g_\sigma({\bf u}-{\bf u}')) \big)\Big] \\
{\mathbb K}^{(HH)}_{\sigma\sigma'}({\bf u},{\bf u}') & = \frac{2}{\epsilon_0+\epsilon_\sigma} \Big[ {\bf n}({\bf u})  \times \big( \cdot \times \nabla ( - \epsilon_0 g_0({\bf u}-{\bf u}') + \delta_{\sigma\sigma'} \epsilon_\sigma  g_\sigma({\bf u}-{\bf u}')) \big)\Big] \\
{\mathbb K}^{(EH)}_{\sigma\sigma'}({\bf u},{\bf u}') & = \frac{2}{\mu_0+\mu_\sigma} \Big[ \kappa (-\epsilon_0\mu_0  g_0({\bf u}-{\bf u}') +\delta_{\sigma\sigma'}\epsilon_\sigma\mu_\sigma  g_\sigma({\bf u}-{\bf u}')){\bf n}({\bf u}) \times \cdot \\
&  + \frac{1}{\kappa}{\bf n}({\bf u}) \times(\cdot \, \nabla) \nabla ( g_0({\bf u}-{\bf u}'))- g_\sigma({\bf u}-{\bf u}')))
\Big]\\
{\mathbb K}^{(HE)}_{\sigma\sigma'}({\bf u},{\bf u}') & = - \frac{\mu_0+\mu_\sigma}{\epsilon_0+\epsilon_\sigma} {\mathbb K}^{(EH)}_{\sigma\sigma'}({\bf u},{\bf u}') \, .
\end{aligned}
\end{equation}
This surface operator $ \mathbb{K}$  has unique mathematical properties which we shall discuss in detail in Sec.~VI. 

The corresponding choice for the coefficients of the integral equations for the surface charges [Eqs.~(\ref{eq:IE_rho_j}), (\ref{eq:IE_rho_m})] are 
\begin{equation}
\label{eq:coeff_c1_static}
\mathbb{c}^{i}_{j,\sigma} = \epsilon_0, \quad \mathbb{c}^{i}_{m,\sigma} = \mu_0, \quad
\mathbb{c}^{e}_{j,\sigma} = \epsilon_\sigma, \quad \mathbb{c}^{e}_{m,\sigma} = \mu_\sigma \, .
\end{equation}

\noindent
{\bf (C2)} An asymmetric, material independent choice of coefficient matrices is 
\begin{equation}
\mathbb{C}^{i}_\sigma={\rm diag}(1,0)\, , \quad \mathbb{C}^{e}_\sigma={\rm diag}(0,1) \, .
\end{equation}
For good conductors, we have observed fast convergence of the MSE with this choice, while   for materials with a moderately high permittivity, like Si,  convergence  is slow, which made us prefer the choice (C1) in the numerical computations in \cite{emig2023}. 
The corresponding expressions of the SSO's ${\mathbb K}$ and of the operator $\mathbb{M}$ are
\begin{equation}
\begin{aligned}
\mathbb{K}^{(EE)}_{\sigma\sigma'}({\bf u},{\bf u}') & =  2 \,{\bf n}_\sigma({\bf u}) \times  \mathbb{G}_0^{(HE)}({\bf u},{\bf u}')   \\
\mathbb{K}^{(HH)}_{\sigma\sigma'}({\bf u},{\bf u}') & =  2 \, \delta_{\sigma \sigma'}\,{\bf n}_\sigma({\bf u}) \times  \mathbb{G}_{\sigma}^{(EH)}({\bf u},{\bf u}')   \\
\mathbb{K}^{(EH)}_{\sigma\sigma'} ({\bf u},{\bf u}') & =  2 \,{\bf n}_\sigma({\bf u}) \times  \mathbb{G}_0^{(HH)}({\bf u},{\bf u}')   \\
\mathbb{K}^{(HE)}_{\sigma\sigma'}({\bf u},{\bf u}') & =  2  \,\delta_{\sigma \sigma'}\,{\bf n}_\sigma({\bf u}) \times  \mathbb{G}_{\sigma}^{(EE)}({\bf u},{\bf u}')  \;,\label{KC2}
\end{aligned}
\end{equation}
and
\begin{equation}
\begin{aligned}
\mathbb{M}^{(EE)}_{\sigma}({\bf u},{\bf r}) & =  2\, {\bf n}_\sigma({\bf u}) \times \mathbb{G}_0^{(HE)}({\bf u},{\bf r})\,, \\
\mathbb{M}^{(EH)}_{\sigma}({\bf u},{\bf r}) & =  2\,{\bf n}_\sigma({\bf u}) \times \mathbb{G}_0^{(HH)}({\bf u},{\bf r})\\
\mathbb{M}^{(HE)}_{\sigma}({\bf u},{\bf r}) & = 0\,, \\
\mathbb{M}^{(HH)}_{\sigma}({\bf u},{\bf r})  &=  0\, .
\end{aligned}
\end{equation}

\noindent
{\bf (C3)} Finally, we note that the singular choice with $\mathbb{C}^{i}_\sigma+\mathbb{C}^{e}_\sigma=\mathbb{0}$, which we excluded, does not yield a Fredholm integral equation and hence does not permit a MSE. A corresponding popular choice  \cite{harrington} is 
\begin{equation}
\mathbb{C}^{i}_\sigma={\rm diag}(-1,-1)\, , \quad \mathbb{C}^{e}_\sigma={\rm diag}(1,1) \, .
\end{equation}
The resulting integral equations for the surface currents are:
\begin{equation}
\sum_{\sigma'=1}^N\int_{S_{\sigma'}}\!\!ds_{{\bf u}'} \,\mathbb{B}_{\sigma\sigma'}({\bf u},{\bf u}')\big(\begin{smallmatrix} {\bf j}_{\sigma'} \\ {\bf m}_{\sigma'} \end{smallmatrix} \big) ({\bf u}')
= \!\int d{\bf r} \,\mathbb{M}_{\sigma}({\bf u},{\bf r})\big(\begin{smallmatrix} {\bf J} \\ {\bf M}\end{smallmatrix}\big) ({\bf r})\;,\label{john}
\end{equation}
with
\begin{equation}
\begin{aligned}
\mathbb{B}_{\sigma\sigma'}({\bf u},{\bf u}') & =   \left[\mathbb{G}_0({\bf u},{\bf u}') + \delta_{\sigma\sigma'} \mathbb{G}_\sigma({\bf u},{\bf u}')\right]_t\\
 \end{aligned} \;,
\end{equation}
\begin{equation}
\begin{aligned}
\mathbb{M}_{\sigma}({\bf u},{\bf r}) & = -  \left[\mathbb{G}_0({\bf u},{\bf r})  \right]_t\\
 \end{aligned} \;,
\end{equation}
where the subscript $t$ means that when  the argument of the tensor  belongs to   the surface $S_{\sigma}$, the tangential projection of the corresponding index of the tensor onto  $S_{\sigma}$ at that position  is  taken.
In \cite{harrington} it is shown that  Eq.(\ref{john})  determines uniquely the surface current at all frequencies. These  integral equations (\ref{john}) been employed in a computationally intensive boundary element method \cite{reid2013}, implemented in the open-source software SCUFF-EM \cite{scuff}.

\section{Limiting Cases}

\subsection{Zero frequency}
\label{sec:static}

The surface integral equations for the currents become singular in the limit of zero-frequency. This singularity does not constitute a problem for evaluation of forces and energies at zero temperature
since both involve integration over all imaginary frequencies. However, it impedes evaluation of the $n=0$ term of the Matsubara sum at finite temperatures. 
Independent of this, one feels that  solving the EM scattering problem at zero frequency in terms of surface currents is somewhat unnatural, and that a simpler approach based solely on surface charges should be possible in the static limit.   We show below that this expectation is indeed correct.

Let us consider the electrostatic problem first.  At points ${\bf r}$ away from the bodies surfaces,  the electrostatic potential $\phi$ satisfies the Equation:
\begin{eqnarray}
{\bf \nabla} \cdot [\epsilon_0{\bf \nabla} \phi ({\bf r})] &=& - \rho_j({\bf r})\;,\;\;\;\;\;\;\;{\bf r} \in   {V_0} \\
{\bf \nabla} \cdot [\epsilon_{\sigma}{\bf \nabla} \phi ({\bf r})] &=&  0\;, \;\;\;\;\;\;\;\;\;\;\;\;\;\;\;\;{\bf r} \in   {V_{\sigma}}\;. \label{lapla2}
\end{eqnarray}
where $ \rho_j$ are the external sources of the incident electrostatic field.
The potential is continuous across the surfaces of the bodies, while its normal derivative  satisfies  the b.c.:
\be
\epsilon_{\sigma}\;\hat {\bf n}_{\sigma} \cdot {\bf \nabla}_- \phi=\epsilon_{0}\;\hat{\bf n}_{\sigma} \cdot {\bf \nabla}_+ \phi\,,\label{bcscal}
\ee 
i.e. the normal component of the induction vector ${\bf D}({\bf r}) =\epsilon({\bf r}) \;{\bf E}({\bf r}) $ is continuous across the surfaces. It is known from potential theory that the scalar potential $\phi$ is determined, within each of the regions $V_0,V_1,\cdots, V_N$ by knowledge of  the external source $\rho_j$ and of the normal derivative of $\phi$, or what is the same by knowledge of the normal component ${\bf D}_n$ of the induction vector, on the surfaces $S_1, \cdots, S_N$. This means that the scattering problem is solved, if we can set up an equation to compute ${\bf D}_n$. To achieve this, we can use a variant of the equivalence principle.  One notes that it is immaterial to replace $\epsilon_{\sigma}$ by $\epsilon_0$ in Eq. (\ref{lapla2}). This means that  away from the surfaces $S_{\sigma}$ the potential  $\phi$ also satisfies Poisson equation  for a {\it homogeneous} medium with permittivity $\epsilon_0$:
\be
 \triangle \phi ({\bf r}) =- \frac{ \rho_j({\bf r})}{\epsilon_0}\;, \label{scal1} 
\ee
When considered in such an  homogeneous medium, the normal component of the corresponding induction vector  ${\bf D}_0=\epsilon_0 \,{\bf E} $ has a jump across the surfaces of the bodies. This discontinuity  of ${\bf D}_0$ can be interpreted as arising from an {\it unphysical} surface distribution of charge ${\bar \varrho}_{\sigma}$ such that:
\be
\bar{{\varrho}}_{j,\sigma}=-{\epsilon_0} \left[ \hat {\bf n}_{\sigma} \cdot {\bf \nabla}_+ \phi -\hat{\bf n}_{\sigma} \cdot {\bf \nabla}_- \phi \right]\;.\label{scal2}
\ee 
In view of Eqs. (\ref{scal1}) and (\ref{scal2}), the potential can be then  expressed {\it everywhere} as:
\be
\phi ({\bf r})=\phi_{\rm inc}({\bf r})+ {\bar \phi}({\bf r})\;,\label{splitsc}
\ee
where
\be
\phi_{\rm inc}  ({\bf r})= \frac{1}{\epsilon_0}\int_{V_0} d^3 {\bf r}' g_0( |{\bf r}-{\bf r}' |)\, \rho_j({\bf r}')\;,\ee
and
\be
{\bar \phi} ({\bf r})=  \frac{1}{\epsilon_0}  \sum_{\sigma=1}^N \int_{S_{\sigma}} d s_{\bf u}\; g_0( |{\bf r}-{\bf u}|)\, \bar{\varrho}_{j,\sigma}({\bf u})\;,\label{barphi}
\ee
with
\be
g_0({\bf r})= \frac{1}{4 \pi |\bf{r}|}\;.
\ee
We note that according to Eq. (\ref{splitsc}), the field  $ {\bar \phi}({\bf r})$ can be identified with the scattered field, at points ${\bf r}$ outside the bodies:
\be
 \phi_{\rm scat}({\bf r})={\bar \phi}({\bf r})\;,\;\;\;\;{\bf r} \in V_0\;.
\ee
An integral Equation for $\bar{\varrho}_{\sigma}$ can be derived as follows. By taking the gradient of Eq. (\ref{splitsc}), one derives the identity:
\be
\hat{\bf n}_{\sigma} \cdot {\bf \nabla} \phi({\bf u})=\hat{\bf n}_{\sigma} \cdot {\bf \nabla} \phi_{\rm inc}({\bf u})+ \hat{\bf n}_{\sigma} \cdot {\bf \nabla} {\bar \phi}({\bf u})\;.\label{step1sc}
\ee
Now, the normal derivative of $\phi$ satisfies the identity:
\be
\hat{\bf n}_{\sigma} \cdot {\bf \nabla} \phi({\bf u})=\frac{1}{2} \left[ \hat{\bf n}_{\sigma} \cdot {\bf \nabla}_+\phi({\bf u}) + \hat{\bf n}_{\sigma} \cdot {\bf \nabla}_- \phi({\bf u})\right]\;.\label{semisum}
\ee
On the other hand, using Eqs. (\ref{bcscal}) and (\ref{scal2}) one finds:
\begin{eqnarray}
\hat{\bf n}_{\sigma} \cdot {\bf \nabla}_+\phi({\bf u}) &=& \frac{\epsilon_{\sigma}}{\epsilon_0}\frac{1}{\epsilon_0-\epsilon_{\sigma}  }  \bar{\varrho}_{j,\sigma} ({\bf u}) \;,\nonumber\\
\hat{\bf n}_{\sigma} \cdot {\bf \nabla}_-\phi({\bf u}) &=&  \frac{1}{ \epsilon_0-\epsilon_{\sigma} }  \bar{\varrho}_{j,\sigma} ({\bf u}) \;.
\end{eqnarray}
Plugging the r.h.s. of the above identities into the r.h.s. of Eq. (\ref{semisum}), we then find:
\be
\hat{\bf n}_{\sigma} \cdot {\bf \nabla}  \phi({\bf u})= \frac{\epsilon_0+\epsilon_{\sigma}}{2\epsilon_0\,( \epsilon_0-\epsilon_{\sigma})} \,\bar{\varrho}_{j,\sigma}({\bf u})\;, \label{nablaphi}
\ee
Upon substituting the r.h.s. of Eq. (\ref{nablaphi}) into the l.h.s. of Eq. (\ref{step1sc}), and expressing the incident field in terms of the external charge $\rho_j$, after a little algebra one obtains a Fredholm integral Equation for $\bar{\varrho}_{\sigma}$. 
The magneto-static problem can be treated in exactly the same way by doing the substitutions $\bar{\varrho}_{j,\sigma} \rightarrow \bar{\varrho}_{m,\sigma}$, $\epsilon_0 \rightarrow  \mu_0$, $\epsilon_{\sigma} \rightarrow \mu_{\sigma}$.
The resulting integral equations for the surface charges  are:
\begin{align}
\label{inteqch:j}
\bar{\varrho}_{j,\sigma}({\bf u}) - \sum_{\sigma'=1}^N \int_{S_{\sigma'}} ds_{{\bf u}'} \, {\mathbb k}^{(j)}_{\sigma\sigma'}({\bf u},{\bf u}') \bar{\varrho}_{j,\sigma'}({\bf u}')  &= \int d{\bf r}\, {\mathbb m}^{(j)}_\sigma({\bf u}, {\bf r})\big(\begin{smallmatrix} \rho_j \\ {\rho_m}\end{smallmatrix}\big) ({\bf r})\;,  \\
\label{inteqch:m}
\bar{\varrho}_{m,\sigma}({\bf u}) - \sum_{\sigma'=1}^N \int_{S_{\sigma'}} ds_{{\bf u}'} \, {\mathbb k}^{(m)}_{\sigma\sigma'}({\bf u},{\bf u}') \bar{\varrho}_{m,\sigma'}({\bf u}')  &= \int d{\bf r}\, {\mathbb m}^{(m)}_\sigma({\bf u}, {\bf r}) \big(\begin{smallmatrix} \rho_j\\ \rho_m\end{smallmatrix}\big) ({\bf r})\;.
\end{align}
Here the kernels are given by
\begin{equation}
\begin{aligned}
\label{eq:small_k_kernel}
{\mathbb k}^{(j)}_{\sigma\sigma'}({\bf u},{\bf u}') &= 2 \, \frac{\epsilon_0-\epsilon_\sigma}{\epsilon_0+\epsilon_\sigma} \, \partial_{{\bf n}_\sigma({\bf u})} \, g_0({\bf u}-{\bf u}')\\
{\mathbb k}^{(m)}_{\sigma\sigma'}({\bf u},{\bf u}') &= 2 \, \frac{\mu_0-\mu_\sigma}{\mu_0+\mu_\sigma} \, \partial_{{\bf n}_\sigma({\bf u})} \, g_0({\bf u}-{\bf u}') \, ,
\end{aligned}
\end{equation}
which turn out to be independent of $\sigma'$, and
\begin{equation}
\begin{aligned}
{\mathbb m}^{(j)}_{\sigma}({\bf u},{\bf r}) &= 2 \, \frac{\epsilon_0-\epsilon_\sigma}{\epsilon_0+\epsilon_\sigma}  \,    \partial_{{\bf n}_\sigma({\bf u})} \, g_0({\bf u}-{\bf r})\\
{\mathbb m}^{(m)}_{\sigma}({\bf u},{\bf r}) &= 2 \, \frac{\mu_0-\mu_\sigma}{\mu_0+\mu_\sigma}  
 \,   \partial_{{\bf n}_\sigma({\bf u})} \, g_0 ({\bf u}-{\bf r})\, .
\end{aligned}
\end{equation}
We note that above integral equations are of the same form as the ones for the surface currents, Eq.~(\ref{eq:Fredholm_currents}).

It is nice to verify that the integral equations  (\ref{inteqch:j}) and  (\ref{inteqch:m}) can be also derived by taking the static limit of  equations (\ref{eq:IE_rho_j}), (\ref{eq:IE_rho_m}), respectively Consider indeed the integral equations (\ref{eq:IE_rho_j}), (\ref{eq:IE_rho_m}) for $\kappa=0$.  
The first term of the integrand $\sim \kappa$ obviously vanishes. In addition, $\lim_{\kappa\to 0}g_\sigma({\bf u}-{\bf u}')=\lim_{\kappa\to 0} g_0({\bf u}-{\bf u}')=1/(4\pi|{\bf u}-{\bf u}'|)$.  
We make now the choice (C1) for the coefficients, see Eq.~(\ref{eq:coeff_c1_static}).
Then, in the static limit, the integral equations read
\begin{align}
\label{eq:rho_static_limit_j}
\varrho_{j,\sigma}({\bf u}) & - \frac{2 \epsilon_\sigma}{\epsilon_0+\epsilon_\sigma} \sum_{\sigma'=1}^N \int_{S_{\sigma'}} ds_{{\bf u}'} \bigg[ \left( (1-\delta_{\sigma\sigma'}) \nabla_{\bf u} g_0({\bf u}-{\bf u}') \times {\bf n}_\sigma ({\bf u})\right){\bf m}_{\sigma'}({\bf u}') \nonumber\\
&- \left(1-\frac{\epsilon_0}{\epsilon_\sigma} \delta_{\sigma\sigma'}\right) \partial_{{\bf n}_\sigma({\bf u})} \varrho_{j,\sigma'}({\bf u}')\bigg] \nonumber\\
&=\frac{2\epsilon_\sigma}{\epsilon_0+\epsilon_\sigma}   {\bf n}_\sigma({\bf u})  {\bf E}_{\rm inc}({\bf u})\\
\label{eq:rho_static_limit_m}
\varrho_{m,\sigma}({\bf u}) & - \frac{2 \mu_\sigma}{\mu_0+\mu_\sigma} \sum_{\sigma'=1}^N \int_{S_{\sigma'}} ds_{{\bf u}'} \bigg[ \left( (\delta_{\sigma\sigma'}-1) \nabla_{\bf u} g_0({\bf u}-{\bf u}') \times {\bf n}_\sigma ({\bf u})\right){\bf j}_{\sigma'}({\bf u}') \nonumber\\
&- \left(1-\frac{\mu_0}{\mu_\sigma} \delta_{\sigma\sigma'}\right) \partial_{{\bf n}_\sigma({\bf u})} \varrho_{m,\sigma'}({\bf u}')\bigg] \nonumber\\
&=\frac{2\mu_\sigma}{\mu_0+\mu_\sigma}   {\bf n}_\sigma({\bf u})  {\bf H}_{\rm inc}({\bf u}) \, .
\end{align}

The term of the sum with $\sigma'=\sigma$ is independent of the surface currents ${\bf j}_{\sigma}$, ${\bf m}_{\sigma}$ due to the delta function. To simplify the terms with $\sigma'\neq\sigma$ we
note that the EM field for ${\bf r}$ located in the interior region of the surface $S_{\sigma'}$ in the static limit can be written as
\begin{align}
{\bf E}^{(\sigma')}({\bf r}) & = \int_{S_{\sigma'}} ds_{\bf u} \left[ \frac{\epsilon_0}{\epsilon_{\sigma'}} \varrho_{j,\sigma'}({\bf u}) \nabla g_0({\bf r}-{\bf u}) + \nabla  g_0({\bf r}-{\bf u}) \times {\bf m}_{\sigma'}({\bf u})\right] \\
{\bf H}^{(\sigma')}({\bf r}) & = \int_{S_{\sigma'}} ds_{\bf u} \left[ \frac{\mu_0}{\mu_{\sigma'}} \varrho_{m,\sigma'}({\bf u}) \nabla g_0({\bf r}-{\bf u}) - \nabla  g_0({\bf r}-{\bf u}) \times {\bf j}_{\sigma'}({\bf u})\right] \, .
\end{align}
If ${\bf r}$ is located in the region exterior to the surface $S_{\sigma'}$, the above integrals vanish. Since  ${\bf u}$ in  Eqs.~(\ref{eq:rho_static_limit_j}), (\ref{eq:rho_static_limit_m}) is located outside of the surface $S_{\sigma'}$ for $\sigma'\neq\sigma$ we can use this relation to eliminate the surface currents.  Upon expressing now $({\varrho}_{j,\sigma}, {\varrho}_{m,\sigma})$ in terms of $(\bar{\varrho}_{j,\sigma},\bar{\varrho}_{m,\sigma})$   via the relations:
\begin{align}
\varrho_{j,\sigma}({\bf u})&=-\frac{\epsilon_{\sigma}}{\epsilon_0} \frac{1}{\epsilon_0- \epsilon_{\sigma}}\,\bar{\varrho}_{j,\sigma}({\bf u})\;,\\
\varrho_{m,\sigma}({\bf u})&=-\frac{\mu_{\sigma}}{\mu_0} \frac{1}{\mu_0- \mu_{\sigma}}\,\bar{\varrho}_{j,\sigma}({\bf u})
\end{align}
which follow from a comparison  of the second of Eqs. (\ref{surchar}) with
Eq. (\ref{nablaphi}) (and the analogous relation for the magnetic field),  and  taking as  incident fields the electrostatic and magnetostatic  fields generated by  external charges $\rho_j$ and $\rho_m$, respectively,  one finds that Eqs. (\ref{eq:rho_static_limit_j}) and (\ref{eq:rho_static_limit_m}) actually coincide with Eqs.  (\ref{inteqch:j}) and (\ref{inteqch:m}), respectively.

For the  benefit of the reader, we write below the expressions of the classical $n=0$ contributions to the Casimir and CP energies, in terms of the static SSO introduced above. They are:
\begin{equation}
\label{eq:4zero}
{\cal E}\vert_{n=0} =\frac{k_B T}{2}  \sum_{p=j,m} {\rm Tr} \log \left[ \mathbb{1} - (\mathbb{1}-\mathbb{k}^{(p)}_{11})^{-1} \mathbb{k}^{(p)}_{12}(\mathbb{1}-\mathbb{k}^{(p)}_{22})^{-1}\mathbb{k}^{(p)}_{21}\right] \, ,
\end{equation}
\begin{equation}
{\cal E}_\text{CP} \vert_{n=0}= - 2\pi k_B T  \,   \sum_{i,j=1}^{3} \left[ \alpha_{ij}(0) \tilde{\mathbb{\Gamma}}^{(EE)}_{ij}({\bf r}_0,{\bf r}_0)+ \beta_{ij}(0) \tilde{\mathbb{\Gamma}}^{(HH)}_{ij}({\bf r}_0,{\bf r}_0) \right] \, ,
\end{equation}
where
\begin{align}
\label{eq:Gammazero}
\tilde{\mathbb{\Gamma}}^{(EE)}({\bf r},{\bf r}')&={\stackrel{\rightarrow}{\bf \nabla}_{\bf r}}
\int_S ds_{\bf u} \int_S ds_{{\bf u}'}\, g_0({\bf r},{\bf u}) (\mathbb{1}-\mathbb{k}^{(j)})^{-1}({\bf u},{\bf u}') \mathbb{m}^{(j)}({\bf u}',{\bf r}')\; {\stackrel{\leftarrow}{\bf \nabla}_{\bf r '}}\;,\\
\tilde{\mathbb{\Gamma}}^{(HH)}({\bf r},{\bf r}')&={\stackrel{\rightarrow}{\bf \nabla}_{\bf r}}
\int_S ds_{\bf u} \int_S ds_{{\bf u}'}\, g_0({\bf r},{\bf u}) (\mathbb{1}-\mathbb{k}^{(m)})^{-1}({\bf u},{\bf u}') \mathbb{m}^{(m)}({\bf u}',{\bf r}')\; {\stackrel{\leftarrow}{\bf \nabla}_{\bf r '}}\;,
\end{align}

\subsection{High frequencies}

It is instructive to study the limit of asymptotically high frequencies. 
We do this here by assuming {\it fixed}, i.e., frequency independent permittivities.
In high frequency limit, the SSO becomes ultra-local, and hence the surface can be approximated by its tangent plane at each position. Then a simple computation yields the following limits 
\begin{align}
\lim_{\kappa \rightarrow \infty} \mathbb{K}^{(EE)}_{\sigma \sigma'} &= \lim_{\kappa \rightarrow \infty} \mathbb{K}^{(HH)}_{\sigma \sigma'}=0 \nonumber \\
\lim_{\kappa \rightarrow \infty} \mathbb{K}^{(EH)}_{\sigma \sigma'} &=  
\begin{pmatrix}
0 & \frac{\sqrt{\epsilon_{0} \mu_{0}}-\sqrt{\epsilon_{\sigma} \mu_{\sigma}} }{\mu_0+\mu_{\sigma}} \\
 \frac{-\sqrt{\epsilon_{0} \mu_{0}}+\sqrt{\epsilon_{\sigma} \mu_{\sigma}} }{\mu_0+\mu_{\sigma}}& 0 \\
\end{pmatrix}\delta_{\sigma\sigma'} \nonumber \\
\lim_{\kappa \rightarrow \infty} 
\mathbb{K}^{(HE)}_{\sigma \sigma'}&=
\begin{pmatrix}
0 & \frac{-\sqrt{\epsilon_{0} \mu_{0}}+\sqrt{\epsilon_{\sigma} \mu_{\sigma}} }{\epsilon_0+\epsilon_{\sigma}} \\
\frac{\sqrt{\epsilon_{0} \mu_{0}}-\sqrt{\epsilon_{\sigma} \mu_{\sigma}} }{\epsilon_0+\epsilon_{\sigma}}& 0 \\
\end{pmatrix}
\, \delta_{\sigma\sigma'}   \, .
\label{liminfU}
\end{align}
Here the matrix elements are expressed in an orthogonal basis of
tangential unit vectors.
This shows that for $\kappa \rightarrow \infty$, the $N$-body $\mathbb{K}$ operator splits into $N$ independent off-diagonal  single body multiplicative operators $\mathbb{K}_{\sigma \sigma}\vert_{\kappa=\infty}$. Using above limits, it is straightforward to verify that the eigenvalues of $\mathbb{K}_{\sigma \sigma}\vert_{\kappa=\infty}$ are 
\be
\lambda_{\sigma; \kappa=\infty}= \pm \frac{ \sqrt{\epsilon_{\sigma} \mu_{\sigma}} - \sqrt{\epsilon_{0} \mu_{0}}}{ \sqrt{(\mu_{\sigma}+\mu_0) (\epsilon_{\sigma} + \epsilon_0)}} \;.\label{eighigh}
\ee
It can be easily verified that for all constant values of the permittivities  $|\lambda_{\sigma;\kappa=\infty}| < 1$, which shows that the MSE converges in the $\kappa \rightarrow \infty$ limit.

\subsection{Perfect conductors}

In the limit of perfect conductors, the boundary conditions reduce to the requirement that the tangential component of the electric field vanish. Hence, it is sufficient to consider only electric surface currents. Those currents are determined by a Fredholm integral equation of the 2nd kind with the operators
\be
\label{eq:K_PC}
\mathbb{K}^{(\rm PC)}_{\sigma\sigma'}({\bf u},{\bf u}')  =  2 \,{\bf n}_\sigma({\bf u}) \times  \mathbb{G}_0^{(HE)}({\bf u},{\bf u}') \;, 
\ee
and
\be
\label{eq:M_PC}
\mathbb{M}^{(\rm PC)}_{\sigma\sigma'}({\bf u},{\bf r})  =  2 \,{\bf n}_\sigma({\bf u}) \times  \mathbb{G}_0^{(HE)}({\bf u},{\bf r}) 
\ee
acting only on the electric surface currents ${\bf j}_{\sigma}({\bf u})  = \hat{\bf n}_{\sigma}({\bf u}) \times {\bf H}_+({\bf u})$.
This result was derived in the study of the Casimir effect for perfectly conducting bodies in  \cite{balian1977}. Details of the derivation of this result are provided in App.~C.

\section{Convergence properties of the MSE} 
\label{sec:conver}

Now we turn to the important problem of the convergence of the Neumann series with the choice (C1) for the coefficients for the SSO ${\mathbb K}$.  It has been shown that the equation $\mathbb{K} {\bf v} = {\bf v}$ does not have any solutions, apart from  the trivial one ${\bf v}={\bf 0}$ \cite{muller}. Since $\mathbb{K}$ is compact, general theorems on compact operators  then ensure that the operator   $({\mathbb I}- \mathbb{K})^{-1} $ exists and is a bounded operator \cite{dunford}. Inversion of the Fredholm integral equation then gives
\be
\big(\begin{smallmatrix} {\bf j} \\ {\bf m}\end{smallmatrix}\big) = ({\mathbb I}- \mathbb{K})^{-1} \,{\mathbb M}\,\big(\begin{smallmatrix} {\bf J} \\ {\bf M}\end{smallmatrix}\big)\;.
\ee
If the Neumann series converges,  $\big(\begin{smallmatrix} {\bf j} \\ {\bf m}\end{smallmatrix}\big)$ can be computed by means of the MSE
\be
\big(\begin{smallmatrix} {\bf j} \\ {\bf m}\end{smallmatrix}\big) =   \left(\sum_{k=0}^{\infty}   \mathbb{K}^k  \right)\,{\mathbb M}\,\big(\begin{smallmatrix} {\bf J} \\ {\bf M}\end{smallmatrix}\big) \;.
\ee
An important question is whether this series converges.  Convergence  is ensured if all eigenvalues of $\mathbb{K}$  are smaller than one in modulus. Unfortunately, a general proof of convergence  does not seem possible. However, we can provide several arguments supporting the conjecture that the Neumann  series indeed converges at all frequencies, and for all passive materials. 
The first argument comes from \cite{balian1977} where it was shown that for an isolated compact perfect conductor with a smooth surface  the eigenvalues of  $\mathbb{K}$  are  smaller than one in modulus.  Below we show that  this conclusion remains true also for any number of perfect conductors. 
In addition, our results below will show explicitly that the Neumann series converges in three distinct limits, namely at all frequencies for perfect conductors, and for magneto-dielectric bodies in the limits of asymptotically large frequencies  and vanishing frequencies.

\subsection{Zero frequency}

We begin by considering the static limit $\kappa \rightarrow 0$. Since in the computation of the Casimir energy of two bodies, one only needs consider the  separate Neumann series of the single-body operators $\mathbb{k}_{11}$ and $\mathbb{k}_{22}$, we here only consider the static limit for a single isolated body. In the static limit, the SSO operator of an isolated body is given by the electric and magnetic kernels $\mathbb{k}^{(j)}_{\sigma\sigma}$, $\mathbb{k}^{(m)}_{\sigma\sigma}$ in Eq.~(\ref{eq:small_k_kernel}). 
We now provide a proof of convergence of the Neumann series 
$(\mathbb{1}-\mathbb{k}^{(j)}_{\sigma\sigma})^{-1}$ and $(\mathbb{1}-\mathbb{k}^{(m)}_{\sigma\sigma})^{-1}$. Convergence is demonstrated by proving that the moduli of the eigenvalues $\lambda$ of  $\mathbb{k}_{\sigma \sigma}$ are smaller than one. Here and the following we drop the index $(j)$ and $(m)$. Let us consider the eigenvalue equation for $\mathbb{k}_{\sigma \sigma}$,
\be
\mathbb{k}_{\sigma \sigma} \; \bar{\varrho}_{\sigma} = \lambda \; \bar{\varrho}_{\sigma}\;.
\ee
Note that the eigenvalues $\lambda$ may be complex, a priori, since   $\mathbb{k}_{\sigma \sigma}$ is not hermitian. Let ${\bar \phi}_{\sigma}$ the field generated by the surface charge distribution $\bar{\varrho}_{\sigma}$,
\be
{\bar \phi}_{\sigma} ({\bf r})=    \int_{S_{\sigma}} d s_{\bf u}\; g_0( {\bf r}-{\bf u})\, \bar{\varrho}_{\sigma}({\bf u})\;,\label{barphi1}
\ee 
The eigenvalue Equation is then equivalent to the  integral equation
\be
 2 \, \frac{\epsilon_0-\epsilon_\sigma}{\epsilon_0+\epsilon_\sigma} \,  \;\hat{\bf n}_{\sigma} \cdot {\bf \nabla} {\bar \phi}_{\sigma}({\bf u})= \lambda\,\bar{\varrho}_{\sigma}({\bf u})\;.\label{eigeq}
\ee
By construction, ${\bar \phi}_{\sigma}$ satisfies Laplace equation at all points away from the surface $S_{\sigma}$,
\be
\triangle {\bar \phi}_{\sigma}=0\;.
\ee 
Moreover, at points on the  surface $S_{\sigma}$ the normal derivative of ${\bar \phi}_{\sigma}$ satisfies the identities
\begin{eqnarray}
\bar{\varrho}_{\sigma}&=& \hat {\bf n}_{\sigma} \cdot {\bf \nabla}_- {\bar \phi}_{\sigma} -\hat{\bf n}_{\sigma} \cdot {\bf \nabla}_+ {\bar \phi}_{\sigma} \;,\\
\hat{\bf n}_{\sigma} \cdot {\bf \nabla} {\bar \phi}_{\sigma}&=&\frac{1}{2} \left[ \hat{\bf n}_{\sigma} \cdot {\bf \nabla}_- {\bar \phi}_{\sigma}+ \hat{\bf n}_{\sigma} \cdot {\bf \nabla}_+ {\bar \phi}_{\sigma}\right]\;.
\end{eqnarray}
From the above identities, we obtain
\be
\hat{\bf n}_{\sigma} \cdot {\bf \nabla} {\bar \phi}_{\sigma}= \hat{\bf n}_{\sigma} \cdot {\bf \nabla}_{\pm} {\bar \phi}_{\sigma}\, \pm  \frac{1}{2}\,\bar{\varrho}_{\sigma}  \;.
\ee
Substitution of the  r.h.s. of this identity into the l.h.s. of Eq. (\ref{eigeq}) gives
\be
 \hat{\bf n}_{\sigma} \cdot {\bf \nabla}_{\pm} {\bar \phi}_{\sigma} =\frac{1}{2}\, \left(\frac{\epsilon_0+\epsilon_{\sigma}}{\epsilon_0-\epsilon_{\sigma}} \,\lambda  \mp 1   \right) \,\varrho_{\sigma}\;.\label{relpm}
\ee
Now, consider the positive-definite integrals ${\cal I}_0$ and ${\cal I}_{\sigma}$ defined by
\begin{eqnarray}
{\cal I}_0&=&\int_{\mathbb{R}^3 - V_{\sigma}} d^3 {\bf r} \; {\bf \nabla } {\bar \phi}^*_{\sigma}({\bf r}) \cdot {\bf \nabla } {\bar \phi}_{\sigma}({\bf r}) \;,\nonumber \\
{\cal I}_{\sigma}&=&\int_{V_{\sigma}} d^3 {\bf r} \; {\bf \nabla } {\bar \phi}^*_{\sigma}({\bf r}) \cdot {\bf \nabla } {\bar \phi}_{\sigma}({\bf r}) \, .
\end{eqnarray}
By using Green's theorem, and then considering the identities in Eq. (\ref{relpm}), one finds that the above integrals become
\begin{eqnarray}
{\cal I}_0&=&-  \int_{S_{\sigma}} d s_{{\bf u}} \; {\bar \phi}^*_{\sigma}({\bf u}) \,\hat{\bf n}_{\sigma} \cdot {\bf \nabla}_+ {\bar \phi}_{\sigma}({\bf u}) = \frac{1}{2}  \left(1 - {\lambda} \, \frac{\epsilon_0+\epsilon_{\sigma}}{\epsilon_0-\epsilon_{\sigma}}  \right)  J_{\sigma} \;,\nonumber\\
{\cal I}_{\sigma}&=&  \int_{S_{\sigma}} d s_{{\bf u}} \; {\bar \phi}^*_{\sigma}({\bf u}) \,\hat{\bf n}_{\sigma} \cdot {\bf \nabla}_- {\bar \phi}_{\sigma}({\bf u}) =    \frac{1}{2}  \left(1 + {\lambda} \, \frac{\epsilon_0+\epsilon_{\sigma}}{\epsilon_0-\epsilon_{\sigma}}  \right)  J_{\sigma} \;, \label{identsca}
 \label{conv2}\end{eqnarray}  
where
\be
J_{\sigma}=\int_{S_{\sigma}} d s_{{\bf u}} \,{\bar \phi}^*_{\sigma}({\bf u}) \varrho_{\sigma}({\bf u}) \, .
\ee
Since ${\cal I}_{\sigma}$ are obviously positive, the integrals $J_{\sigma}$ cannot be zero.   Upon multiplying the first of Eqs. (\ref{conv2}) by the conjugate of the second, we obtain
 \be
 {\cal I}_0 {\cal I}_{\sigma}= \frac{1}{4}\left[ 1- {|\lambda|^2} \left(\frac{\epsilon_0+\epsilon_{\sigma}}{\epsilon_0-\epsilon_{\sigma}} \right)^2 - 2\, {\rm i}\; \frac{\epsilon_0+\epsilon_{\sigma}}{\epsilon_0-\epsilon_{\sigma}} \, {\rm Im} \,\lambda \right] |J_{\sigma}|^2\;.
 \ee
 This identity  implies that  ${\rm Im} \,\lambda$=0, and one obtains the inequality
 \be
  1- {|\lambda|^2} \left(\frac{\epsilon_0+\epsilon_{\sigma}}{\epsilon_0-\epsilon_{\sigma}} \right)^2 > 0\;,
\ee 
which directly implies
\be
|\lambda|^2 <  \left(\frac{\epsilon_0-\epsilon_{\sigma}}{\epsilon_0+\epsilon_{\sigma}}\right)^2 < 1\;,
\ee
since $\epsilon_0$ and $\epsilon_{\sigma}$ are both positive numbers. An analogous proof shows that $|\lambda| < 1$ for the magneto-static problem.

\subsection{Perfect conductors}

In this subsection we  prove  that the Neumann series   for a collection of perfectly conducting bodies converges for all imaginary frequencies. The proof applies to  compact bodies, with  smooth surfaces. Convergence is demonstrated by proving that the absolute values  of the eigenvalues $\lambda$ of the operator $\mathbb{K}^{(\rm PC)}$ in Eq. (\ref{eq:K_PC}) are less than one.     We note that convergence of the Neumann series was proved in \cite{balian1977} for a single body, in a larger domain of complex frequencies $\omega$, that includes the imaginary axis.  Since for purely imaginary frequencies the proof becomes considerably simpler, we find it useful to present it here for a general system of  $N$ conductors. Let us consider the eigenvalue equation for $\mathbb{K}^{(\rm PC)}$ 
\be
\mathbb{K}^{(\rm PC)}\;  {\bf j} = \lambda \;  {\bf j}  \;.
\ee
Note that the eigenvalues $\lambda$ may be complex, a priori, since   $\mathbb{K}$ is not hermitean. Let $({\bf E},{\bf H})$ the EM field generated by the surface current ${\bf j}$, 
\begin{eqnarray}
{\bf E}({\bf r})  &=&  
 \sum_{\sigma=1}^N \int_{S_\sigma} d s_{{\bf u}}\, \mathbb{G}_0^{(EE)} ({\bf r}-{\bf u})  {\bf j}_{\sigma}  ({\bf u}) \;,\nonumber \\
{\bf H}({\bf r})  &=&  
  \sum_{\sigma=1}^N \int_{S_\sigma} d s_{{\bf u}}\, \mathbb{G}_0^{(HE)} ({\bf r}-{\bf u})  {\bf j}_{\sigma}  ({\bf u}) \;.\label{Esigma}
\end{eqnarray} 
In view of Eq.~(\ref{eq:K_PC}), we see that the eigenvalue equation is equivalent to the  relation
\begin{eqnarray}
2\, \hat{\bf n}_{\sigma}({\bf u})   \times {\bf H} ({\bf u}) &=& \lambda \, {\bf j}_{\sigma} ({\bf u}) \;.\label{eigEM}
\end{eqnarray}
By construction, the EM field $({\bf E},{\bf H})$ satisfies Maxwell Equations at points ${\bf r}$ not lying on any of the surfaces $S_{\sigma}$, 
\begin{eqnarray}
-{\bf \nabla} \times {\bf E}({\bf r})&=& \kappa \,\mu_0 \;{\bf H}({\bf r}) \;,\\
{\bf \nabla} \times {\bf H}({\bf r}) &=& \kappa \,\epsilon_0\; {\bf E}({\bf r})\;.
\end{eqnarray}
 At points ${\bf u}$ on $S_{\sigma}$ the field $({\bf E} ,{\bf H})$  satisfies the jump conditions:
\begin{eqnarray}
{\hat {\bf n}}_{\sigma}({\bf u}) \times  \left[{\bf E}_+({\bf u})  -{\bf E}_-({\bf u})  \right]&=& 0\;,  \nonumber \\
{\hat {\bf n}}_{\sigma}({\bf u}) \times  \left[{\bf H}_+({\bf u})  -{\bf H}_-({\bf u})  \right]&=&  {\bf j}_{\sigma}\;. \label{jumpEM1}
\end{eqnarray}
Moreover, it holds
\begin{eqnarray}
 {\hat {\bf n}}_{\sigma}({\bf u}) \times  {\bf E}({\bf u})   &=&  \frac{1}{2} \;{\hat {\bf n}}_{\sigma}({\bf u}) \times  \left[{\bf E}_+({\bf u})  + {\bf E}_-({\bf u})  \right] \;,  \nonumber \\
 {\hat {\bf n}}_{\sigma}({\bf u}) \times  {\bf H}({\bf u})   &=&\frac{1}{2} \; {\hat {\bf n}}_{\sigma}({\bf u}) \times  \left[{\bf H}_+({\bf u})  + {\bf H}_-({\bf u})  \right]\;. \label{jumpEM2}
\end{eqnarray}
Combining Eqs.~(\ref{jumpEM1}) and (\ref{jumpEM2}) we obtain
\begin{eqnarray}
{\hat {\bf n}}_{\sigma}({\bf u}) \times  {\bf E}({\bf u}) &=& {\hat {\bf n}}_{\sigma}({\bf u}) \times  {\bf E}_{\pm}({\bf u})  \nonumber \\
{\hat {\bf n}}_{\sigma}({\bf u}) \times  {\bf H}({\bf u}) &=&  {\hat {\bf n}}_{\sigma}({\bf u}) \times  {\bf H}_{\pm}({\bf u})  \mp \frac{1}{2} \;{\bf j}_{\sigma}\;.\label{relfields}
\end{eqnarray}
Upon substituting the r.h.s.~of the second of the above Equations into the l.h.s.~of the eigenvalue Eq.~(\ref{eigEM}), we obtain the  relation
\begin{eqnarray}
 2 \, {\hat {\bf n}}_{\sigma}({\bf u}) \times  {\bf H}_{\pm}({\bf u})&=&  (\lambda \pm 1) \,{\bf j}_{\sigma}  \;.\label{relpmEM}
\end{eqnarray}
Now, consider the  energy fluxes  across the inner and the outer sides of the surface $S_{\sigma}$, given by surface integrals of the Poynting vector,
\be
{\cal J}_{\sigma \pm}=   2\, \int_{S_{\sigma}} d s_{{\bf u}}\, {\hat {\bf n}}_{\sigma}({\bf u}) \cdot \left( {\bf E}^{*}_{\pm} \times {\bf H}_{\pm} \right) ({\bf u})\;.\label{calJ}
\ee
By using the divergence theorem, one  obtains the  identities
\begin{eqnarray}
{\cal J}_{\sigma-} &=& {\cal I}_{\sigma-} \;,\nonumber \\
  \sum_{\sigma=1}^N {\cal J}_{\sigma+} &=&-{\cal I}_{+} \;,\label{iddupl}
\end{eqnarray}
where ${\cal I}_{\sigma-}$ and ${\cal I}_{+}$  denote the  following positive-definite integrals
\begin{eqnarray}
{\cal I}_+&=& 2 \kappa \int_{V_0} d^3 {\bf r} \; \left( \epsilon_0 \,{\bf E}^* \cdot {\bf E} +  \mu_0 \,{\bf H}^* \cdot {\bf H}\right) \;,\nonumber \\
{\cal I}_{\sigma -}&=& 2 \kappa \int_{V_{\sigma}} d^3 {\bf r} \; \left( \epsilon_0\,{\bf E}^{*} \cdot {\bf E} +  \mu_0\,{\bf H}^{*} \cdot {\bf H}\right)\;.
\end{eqnarray}

Upon substituting  Eqs.~(\ref{relpmEM}) into the r.h.s.~of Eq.~(\ref{calJ}), and recalling the first of Eqs.~(\ref{relfields}) we find that  the  identities in Eq.~(\ref{iddupl}) can be recast as
\begin{eqnarray}
{\cal I}_{+} &=&  (1+\lambda)\,    \sum_{\sigma=1}^N \int_{S_{\sigma}} d s_{{\bf u}}\,   {\bf E}^*\cdot {\bf  j}_{\sigma} \;,  \nonumber \\
 {\cal I}_{\sigma-} &=&  (1-\lambda)\,   \int_{S_{\sigma}} d s_{{\bf u}}\,   {\bf E}^* \cdot {\bf  j}_{\sigma}\;.
\end{eqnarray}
 Since ${\cal I}_{+}$ and ${\cal I}_{\sigma-}$ are  positive, neither of the surface integrals on the r.h.s.~of the above equations can be zero. Upon  adding the identities in the second line of the above equation, and then dividing the sum by the  identity in the first line, we find
 \be
 \frac{1-\lambda}{1+\lambda}=\frac{{\cal I}_-}{{\cal I}_+}\;,
 \ee
 where we set ${\cal I}_-= \sum_{\sigma=1}^N {\cal I}_{\sigma -}$. By solving for $\lambda$, we get
 \be
 \lambda=\frac{ {\cal I}_+ - {\cal I}_- }{ {\cal I}_+ + {\cal I}_-}\;.
 \ee
 This relation shows that the eigenvalues are real, and that $|\lambda| < 1$ since ${\cal I}_{+}$, ${\cal I}_{-} >0$.
 This establishes convergence of the Neumann series.

\subsection{Some general properties of the SSO in the formulation (C1)}

In this Section, we derive the main properties of the SSO $\mathbb{K}$ with the coefficient choice (C1) for magneto-dielectric bodies. We assume throughout that the frequency $\omega$ is imaginary $\omega= {\rm i} \xi$, with $\xi>0$. We underline though that most of the properties discussed below are in fact valid for arbitrary frequencies $\omega$ belonging to the upper complex plane  ${\cal C}^+=\{\omega: {\rm Im}(\omega) \ge 0\}$, as the reader may easily verify in each case.   We recall that along the positive imaginary frequency axis the permittivities  of dissipative and dispersive media are positive numbers, and therefore we assume below  $\epsilon_{\sigma} >0$ and $\mu_{\sigma}>0$.  

The unique feature  of the formulation (C1), which distinguishes it from all other formulations,  is its weak short-distance singularity, since $\mathbb{K}$ behaves as $|{\bf u}-{\bf u}'|^{-1}$ when ${\bf u} \rightarrow {\bf u}'$.  We note that an analogous weak singularity is also displayed by the  SSO $\mathbb{K}^{(\rm PC)}$ for perfect conductors in Eq. (\ref{eq:K_PC}). This has to be contrasted with the  $|{\bf u}-{\bf u}'|^{-3}$ singularity displayed by $\mathbb{K}$, for all other choices of the coefficients.  As a result of its weak singularity, the SSO $\mathbb{K}$ is a compact operator \cite{muller}. As it is well known \cite{dunford}, the spectrum $\sigma(\mathbb{A})$ of a compact operator $\mathbb{A}$ consists only of discrete eigenvalues,  and the set of its non-vanishing eigenvalues (each counted as many times as its multiplicity) is either empty, or finite or it is a sequence converging to zero. The latter property implies  that the number of eigenvalues whose modulus exceeds any positive constant is necessarily finite.   An important consequence of this general property of compact operators  is  that the number of eigenvalues of  $\mathbb{K}$ that exceed one in modulus is finite, which implies that the MSE of $({\mathbb I}- \mathbb{K})^{-1}$ converges in general, except possibly in a finite-dimensional subspace.  

Before we study some mathematical properties of the operator $\mathbb{K}$, it is instructive to consider its general structure and behavior of low and high imaginary frequencies $\kappa$.
Consider the expression for $\mathbb{K}$ given in Eq.~(\ref{eq:K_explicit}). For small $\kappa$ the $EH$ and $HE$ components vanish, as can be seen by expanding  $g_\sigma({\bf u}-{\bf u}')$ for small $\kappa$. In the opposite limit of large $\kappa$, the $EE$ and $HH$ components of $\mathbb{K}$ vanish, as we had seen explicitly already in Eq.~(\ref{liminfU}). We shown before already that in both limits the eigenvalues of $\mathbb{K}$ are smaller than one. This implies that the MSE must converge for sufficiently small and for sufficiently large $\kappa$. However, this does not guarantee convergence for all values of $\kappa$ since the eigenvalues are not 
monotonous functions of $\kappa$, as we shall see in the examples given in the next section.

We proceed with some mathematical properties of $\mathbb{K}$. On the space of surface currents 
$\big(\begin{smallmatrix} {\bf j} \\ {\bf m}\end{smallmatrix}\big)$ we define
the scalar product 
\be
\langle \,\big(\begin{smallmatrix} {\bf j'} \\ {\bf m'}\end{smallmatrix}\big) \,|\, \big(\begin{smallmatrix} {\bf j} \\ {\bf m}\end{smallmatrix}\big)\, \rangle =\sum_{\sigma=1}^N \int_{S_{\sigma}} ds_{\bf u} [ \,{\bf j}_{\sigma}'^*({\bf u}) \cdot {\bf j}_{\sigma}({\bf u}) + {\bf m}_{\sigma}'^* ({\bf u})\cdot {\bf m}_{\sigma}({\bf u}) ]\;.
\ee
It is a simple matter to verify  that the $\mathbb{K}$ operator in Eq. (\ref{eq:12}) can be factorized as
\be
\mathbb{K} = \mathbb{R}\,\mathbb{U}  \;, \label{factK}
\ee 
where $\mathbb{R}$ is the local multiplicative operator
\be
\mathbb{R}  \left(\begin{matrix} {\bf j}_{\sigma} \\ {\bf m}_{\sigma} \end{matrix} \right) ({\bf u})=  \left(\begin{matrix}  \hat {\bf n}_{\sigma} ({\bf u}) \times {\bf j}_{\sigma} ({\bf u})\\ - \hat {\bf n}_{\sigma} ({\bf u}) \times {\bf m}_{\sigma} ({\bf u})  \end{matrix} \right) \;,
\ee
and $\mathbb{U}$ is the surface operator
\begin{equation}
\begin{aligned}
\label{Uop}
\mathbb{U}^{(EE)}_{\sigma\sigma'}({\bf u},{\bf u}') & = \frac{2}{\mu_0+\mu_\sigma}   \left[ \mu_0 \mathbb{G}_0^{(HE)}({\bf u},{\bf u}') - \delta_{\sigma\sigma'} \mu_\sigma \mathbb{G}_\sigma^{(HE)}({\bf u},{\bf u}')\right]_t\\
\mathbb{U}^{(HH)}_{\sigma\sigma'}({\bf u},{\bf u}') & = \frac{2}{\epsilon_0+\epsilon_\sigma}  \left[ \epsilon_0 \mathbb{G}_0^{(EH)}({\bf u},{\bf u}') - \delta_{\sigma\sigma'} \epsilon_\sigma \mathbb{G}_\sigma^{(EH)}({\bf u},{\bf u}')\right]_t \\
\mathbb{U}^{(EH)}_{\sigma\sigma'}({\bf u},{\bf u}') & = \frac{2}{\mu_0+\mu_\sigma}  \left[ \mu_0 \mathbb{G}_0^{(HH)}({\bf u},{\bf u}') - \delta_{\sigma\sigma'} \mu_\sigma \mathbb{G}_\sigma^{(HH)}({\bf u},{\bf u}')\right]_t \\
\mathbb{U}^{(HE)}_{\sigma\sigma'}({\bf u},{\bf u}') & = \frac{2}{\epsilon_0+\epsilon_\sigma}   \left[ \epsilon_0 \mathbb{G}_0^{(EE)}({\bf u},{\bf u}') - \delta_{\sigma\sigma'} \epsilon_\sigma \mathbb{G}_\sigma^{(EE)}({\bf u},{\bf u}')\right]_t
\end{aligned} \;,
\end{equation}
where the subscript $t$ denotes projection of tensors onto the tangent plane at $S_{\sigma}$. We note that both $\mathbb{R}$ and $\mathbb{U}$ are real operators.  Let us define the transpose  $\mathbb{A}^{\rm T}$  of an operator $\mathbb{A}$,
 \be
(\mathbb{A}^{\rm T})^{(\alpha \beta )}_{ij;\sigma\sigma'}({\bf u},{\bf u}')=  \mathbb{A}^{(\beta \alpha)}_{ji;\sigma'\sigma}({\bf u}',{\bf u})\;.
\ee
The operator $\mathbb{R}$ is orthogonal,
\be
\mathbb{R} \;\mathbb{R}^{\rm T} = -\mathbb{R}^2 = \mathbb{1}\;.
\ee
It can be verified that $\mathbb{U}$ satisfies the relation
\be
\mathbb{g} \,\mathbb{U} =\mathbb{U}^T  \,\mathbb{g}\;, \label{symU}
\ee
where $\mathbb{g} $ is the local positive and symmetric operator
\be
\mathbb{g} \left(\begin{matrix} {\bf j}_{\sigma} \\ {\bf m}_{\sigma} \end{matrix} \right) ({\bf u})= 
\left(\begin{matrix} (\mu_{\sigma}+\mu_0) \;{\bf j}_{\sigma} \\ (\epsilon_{\sigma}+\epsilon_0)\; {\bf m}_{\sigma} \end{matrix} \right) ({\bf u}) \;.
\ee
We note also that  $\mathbb{R} $ and $\mathbb{g}$ commute,
\be
 [ \,\mathbb{R} , \mathbb{g} \,]=0\;.
\ee 
The symmetry property Eq.~(\ref{symU}) implies that the operator $\mathbb{U}$ is self-adjoint with respect to the following material-dependent inner product $\langle  \; | \; \rangle_g$,
\be
\langle \,\big(\begin{smallmatrix} {\bf j'} \\ {\bf m'}\end{smallmatrix}\big) \,|\, \big(\begin{smallmatrix} {\bf j} \\ {\bf m}\end{smallmatrix}\big)\, \rangle_g \equiv \langle \,\big(\begin{smallmatrix} {\bf j'} \\ {\bf m'}\end{smallmatrix}\big) \,|\,\mathbb{g}\; \big(\begin{smallmatrix} {\bf j} \\ {\bf m}\end{smallmatrix}\big)\, \rangle =\sum_{\sigma=1}^N \int_{S_{\sigma}} ds_{\bf u} [(\mu_{\sigma}+\mu_0) \,{\bf j}_{\sigma}'^*({\bf u}) \cdot {\bf j}_{\sigma}({\bf u}) + (\epsilon_{\sigma}+\epsilon_0) \,{\bf m}_{\sigma}'^* ({\bf u})\cdot {\bf m}_{\sigma}({\bf u}) ]\;.
\ee
Thus,   Eq.~(\ref{factK})  shows that  for imaginary frequencies the operator $\mathbb{K}$ is the product of an orthogonal operator $\mathbb{R}$  times a self-adjoint real operator $\mathbb{U}$. This implies that if $\lambda$ is an eigenvalue, then also $-\lambda$, $\lambda^*$ and $-\lambda^*$ are eigenvalues. We note first that reality of $\mathbb{K}$ implies that the set of its eigenvalues  is formed by pairs $(\lambda , \lambda^*)$ of complex conjugate eigenvalues.  Consider now an eigenvalue $\lambda$ of $\mathbb{K}$. Since the eigenvalues of an operator coincide with the eigenvalues of its transpose, there must exist a non-vanishing left eigenvector $v$ of $\mathbb{K}$ such that
\be
\mathbb{K}^{\rm T} v = -\mathbb{U}^{\rm T} \mathbb{R}\; v = \lambda \,v\;.
\ee
Now we define $w= \mathbb{g}^{-1} \mathbb{R} \;v$.  The vector $w$ is clearly different from zero, because $\mathbb{R}$ is orthogonal and $\mathbb{g}$ is a positive operator.  Then, using the relation Eq.~(\ref{symU}), we get
\be
\mathbb{K} \;w = \mathbb{R} \,\mathbb{U} \, \mathbb{g}^{-1} \mathbb{R} \;v = \mathbb{R} \, \mathbb{g}^{-1} \,\mathbb{U}^{\rm T} \mathbb{R} \;v = -\lambda \mathbb{R} \, \mathbb{g}^{-1}  \,v =  -\lambda\, \mathbb{g}^{-1} \, \mathbb{R}  \,v= -\lambda w \;, 
\ee
which shows that $w$ is an eigenvector of $\mathbb{K}$ with eigenvalue $-\lambda$. It is clear that all the above conclusions are true also for the SSO $\mathbb{K}_{\sigma \sigma}$ of the $\sigma$-th body in isolation.

\section{Examples}

In order to strengthen the case for convergence of the Neumann series for the formulation (C1), we consider in this section explicitly the operator  $\mathbb{K}$ for a magneto-dielectric plate, sphere and cylinder. The eigenvalues can be computed exactly in these cases, using plane waves or the partial-wave representations of the free Green tensors. We considered several distinct values of the electric and magnetic permittivities, and always found that the moduli of the eigenvalues are less than one, at all frequencies.

\subsection{Example 1: Magneto-dielectric parallel plates}

\begin{figure}
\includegraphics[width=0.6\textwidth]{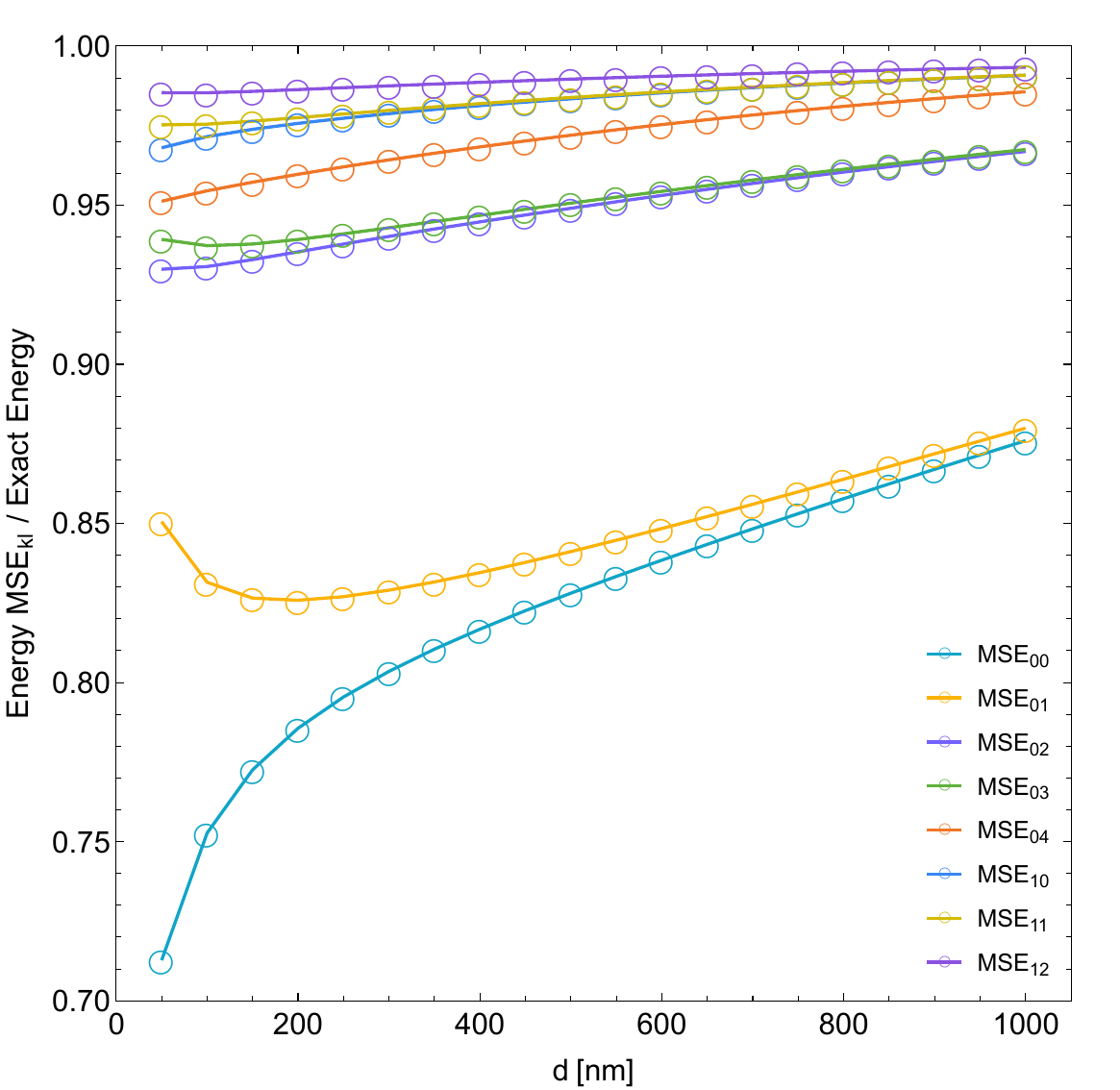} 
\caption{\label{fig:2}
{\bf Multiple scattering expansion of the Casimir energy between a silicon plate and a gold plate.} Different orders of the MSE for the Casimir energy between a plate made of doped silicon and a plate made of gold, normalized to the known exact energy. Indices of MSE$_{kl}$ label the number of scatterings between the plates ($2(k+1)$) and within the silicon plate ($l$) (see text for details).
}
\end{figure}

The Casimir interaction between two planar and parallel surfaces is determined by their Fresnel coefficients according to the Lifshitz formula \cite{lifshitz}. In our formulation, the Casimir interaction is determined by the SSO $\mathbb{K}$. For an infinite, planar surface of a material with permittivities $\epsilon_\sigma$ and $\mu_\sigma$ in an external medium with permittivities $\epsilon_0$ and $\mu_0$, the SSO  $\mathbb{K}_{\sigma\sigma}$ can be expressed easily in a plane wave basis,
\begin{equation}
\begin{aligned}
\mathbb{K}^{(EE)}_{\sigma\sigma} & = 0 \, , \\
\mathbb{K}^{(HH)}_{\sigma\sigma} & = 0 \, , \\
\mathbb{K}^{(EH)}_{\sigma\sigma} & = (-1)^\sigma (2\pi)^2 \delta({\bf k}_\|-{\bf k}'_\|) \frac{1}{\kappa}\frac{1}{\mu_0+\mu_\sigma} \Bigg[ \frac{1}{\sqrt{\epsilon_0\mu_0 \kappa^2+{\bf k}_\|^2}} \begin{pmatrix} -k_1 k_2 & -\epsilon_0\mu_0 \kappa^2-k_1^2 \\  \epsilon_0\mu_0 \kappa^2+k_2^2  & k_1 k_2\end{pmatrix} \\
& - \frac{1}{\sqrt{\epsilon_\sigma\mu_\sigma \kappa^2+{\bf k}_\|^2}} \begin{pmatrix} -k_1 k_2 & -\epsilon_\sigma\mu_\sigma \kappa^2-k_1^2 \\  \epsilon_\sigma\mu_\sigma \kappa^2+k_2^2  & k_1 k_2\end{pmatrix}
\Bigg]\, , \\
\mathbb{K}^{(HE)}_{\sigma\sigma} & = - \frac{\mu_0+\mu_\sigma}{\epsilon_0+\epsilon_\sigma} \mathbb{K}^{(EH)}_{\sigma\sigma}  \, ,
\end{aligned}
\end{equation}
where ${\bf k}_\|=(k_1,k_2)$ is the k-vector parallel to the surface. The factor $(-1)^\sigma$ accounts for the different orientation of the surface normal vector on the two plates. The eigenvalues of  $\mathbb{K}_{\sigma\sigma}$ are
\begin{equation}
\lambda_{\pm}({\bf k}_\|) = \pm \left[ \frac{(s_1-s_0)(\epsilon_1 \mu_1 s_0-\epsilon_0\mu_0 s_1)}{s_0 s_1(\epsilon_0+\epsilon_1)(\mu_0+\mu_1)}\right]^{1/2}
\end{equation}
with $s_\sigma=\sqrt{\epsilon_\sigma \mu_\sigma \kappa^2+{\bf k}_\|^2}$. Each eigenvalue has an algebraic multiplicity of two. The eigenvalues are real valued, and $|\lambda_{\pm}({\bf k}_\|)|<1$ as can be easily checked.

The components of the operators $\mathbb{K}_{\sigma\sigma'}$ with $\sigma\neq\sigma'$ which couple surface currents on different surfaces are also easily expressed in plane waves, leading to
\begin{equation}
\begin{aligned}
\mathbb{K}^{(EE)}_{12} & =  (2\pi)^2 \delta({\bf k}_\|-{\bf k}'_\|) \frac{\mu_0}{\mu_0+\mu_1} \begin{pmatrix} -1 & 0\\ 0 & -1\end{pmatrix} e^{-s_0 d}\\
\mathbb{K}^{(HH)}_{12} & =  (2\pi)^2 \delta({\bf k}_\|-{\bf k}'_\|) \frac{\epsilon_0}{\epsilon_0+\epsilon_1} \begin{pmatrix} -1 & 0\\ 0 & -1\end{pmatrix} e^{-s_0 d}\\
\mathbb{K}^{(EH)}_{12} & =  (2\pi)^2 \delta({\bf k}_\|-{\bf k}'_\|) \frac{1}{(\mu_0+\mu_1)\kappa s_0}\begin{pmatrix} -k_1 k_2 & -k_1^2-\epsilon_0\mu_0 \kappa^2\\ k_2^2+\epsilon_0\mu_0 \kappa^2 & k_1 k_2\end{pmatrix} e^{-s_0 d}\\
\mathbb{K}^{(HE)}_{12} & =  (2\pi)^2 \delta({\bf k}_\|-{\bf k}'_\|) \frac{1}{(\epsilon_0+\epsilon_1)\kappa s_0}\begin{pmatrix} k_1 k_2 & k_1^2+\epsilon_0\mu_0 \kappa^2\\ -k_2^2-\epsilon_0\mu_0 \kappa^2 & -k_1 k_2\end{pmatrix} e^{-s_0 d} \, .
\end{aligned}
\end{equation}
The elements of ${\mathbb K}_{21}$ are obtained from those of ${\mathbb K}_{12}$ by replacing $\epsilon_1$, $\mu_1$ by $\epsilon_2$, $\mu_2$ and changing the sign of the $EH$ and $HE$ components.
When these operator components are substituted into Eq.~(\ref{eq:4}), the Lifshitz formula \cite{lifshitz} is recovered. We note that the inverse of $\mathbb{1}-\mathbb{K}_{\sigma\sigma}$ can be computed easily as the operator is diagonal. However, to examine the convergence rate of the MSE, we expand $(\mathbb{1}-\mathbb{K}_{\sigma\sigma})^{-1}$ into a Neumann series in $\mathbb{K}_{\sigma\sigma}$ and compute the Casimir energy at different orders of the MSE. 
Since $|\lambda_{\pm}({\bf k}_\|)|<1$, the MSE must converge. 
Indeed, when the SSO $\mathbb{K}_{11}$ describes the scatterings on one plate, expansion of the energy in Eq.~(\ref{eq:3}) in this SSO yields MSE approximants to the Casimir interaction. MSE orders are labelled by MSE$_{kl}$ where $2(k+1)$ is the number of scatterings between the surfaces (total number of $\mathbb{K}_{12}$ and $\mathbb{K}_{21}$ operators) and $l$ is the number of single-body scatterings on the Si surface  (number of $\mathbb{K}_{11}$ operators). 

The majority of experiments measure forces between gold (Au) and/or doped silicon (Si) surfaces \cite{Wang2021,mohideen,lamoreaux,bressi}, and hence we consider these materials in this example. 
Figure \ref{fig:2} shows the energy for eight different orders of MSE relative to the known exact energy at $T=300$K for surface separations between 100nm and 1$\mu$m. While the lowest order MSE$_{00}$ with {\it no} single-body scattering on the Si surface yields already between $70\%$ and $87\%$ of the exact interaction, only 4 scatterings between the surfaces ($k=1$) and 2 single-body scatterings on the Si surface ($l=2$) are required for an accuracy of about $1\%$. This validation example  demonstrates fast convergence of our MSE, with good homogeneity in separation.

\subsection{Example 2: Magneto-dielectric sphere}

\begin{figure}
\includegraphics[width=0.75\textwidth]{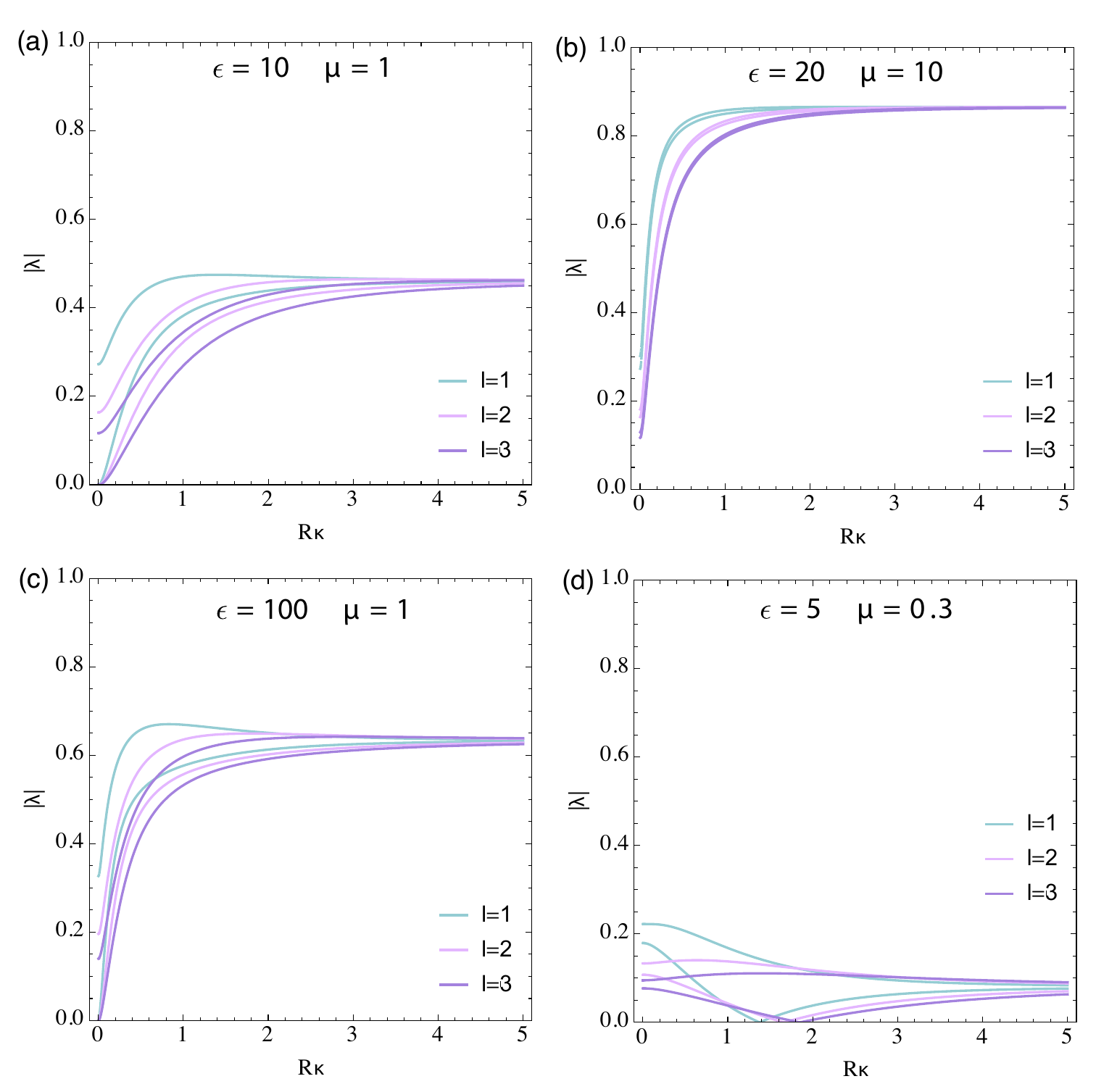} 
\caption{\label{fig:sphere}
{\bf Eigenvalues of $\mathbb{K}_{11}$ for a magneto-dielectric sphere of radius $R$.}
Shown are the absolute values of the eigenvalues for electric and magnetic polarization, as a function of the re-scaled frequency $R\kappa$ for different partial wave indices $l=1,2,3$. For each value of $l$ only two of the four eigenvalues are shown, as eigenvalues of $\mathbb{K}_{11}$ always appear in pairs $(\lambda,-\lambda)$. The permittivities are indicated in the plots.}
\end{figure}

For a magneto-dielectric sphere of radius $R$ the SSO operator $\mathbb{K}_{11}$ can be computed easily in terms of vector spherical harmonics. A similar computation has been carried out in \cite{balian1978} for a perfectly conducting sphere. The elements of the infinite matrix representing $\mathbb{K}_{11}$ can be expressed in terms of Bessel functions $I_{l+1/2}(z)$, $K_{l+1/2}(z)$ with half-integer index. The full expressions are not particularly illuminating and hence are not shown here. 
Fig.~\ref{fig:sphere} shows the eigenvalues of $\mathbb{K}_{11}$ for the first three partial waves as a function of the re-scaled frequency $\kappa R$.  
For all considered permittivities and frequencies, the moduli of the eigenvalues were found to be less than one. For large $R\kappa$ the eigenvalues become independent of the partial wave index $l$ as they approach the high energy limit given by Eq.~(\ref{eighigh}).


\subsection{Example 3:  Magneto-dielectric  cylinder}

\begin{figure}[h]
\includegraphics[width=1.\textwidth]{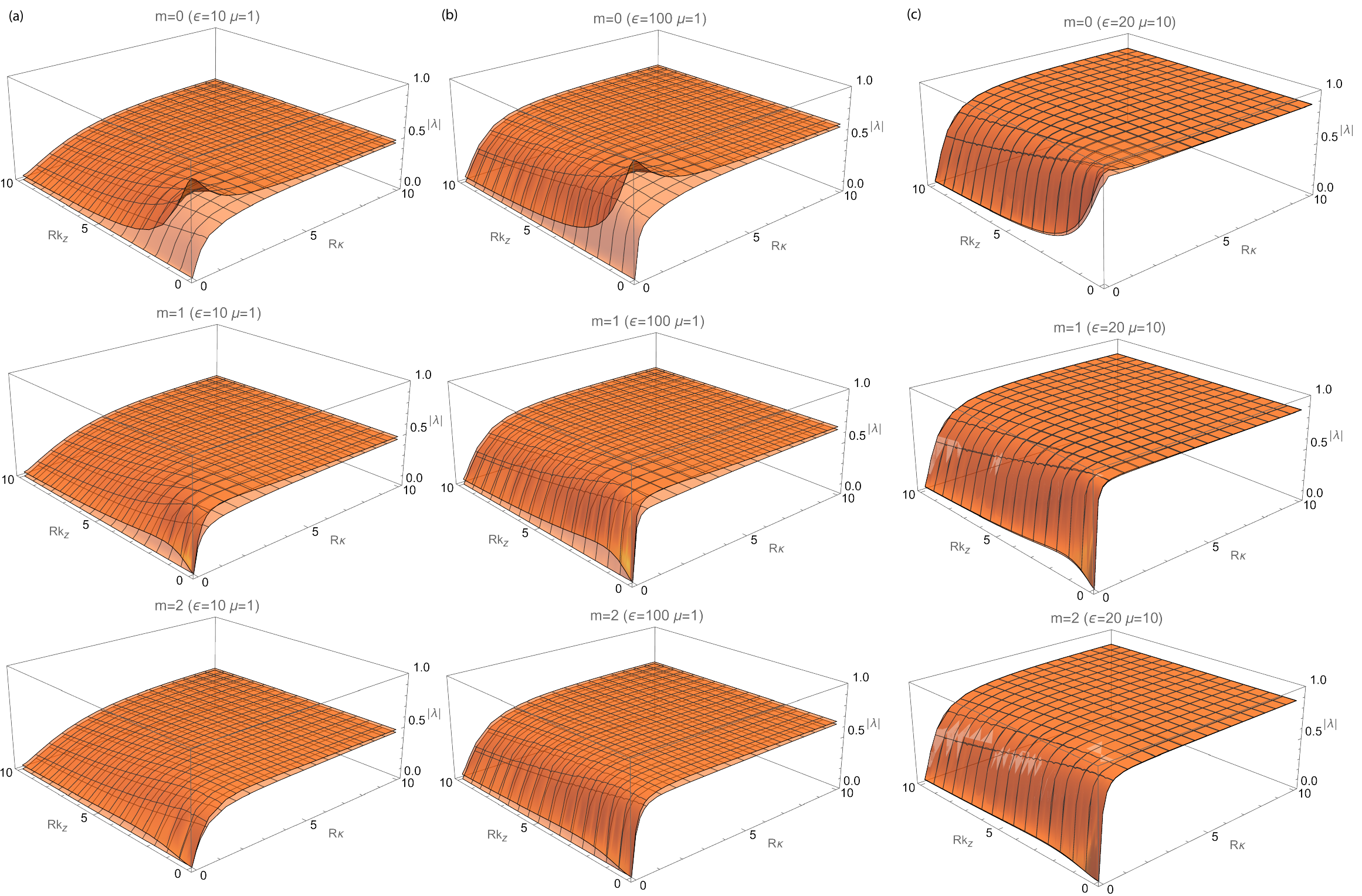} 
\caption{\label{fig:cylinder}
{\bf Eigenvalues of $\mathbb{K}_{11}$ for a magneto-dielectric cylinder of radius $R$.}
Shown are the absolute values of the eigenvalues for electric and magnetic polarization, as a function of the re-scaled frequency $R\kappa$ and the re-scaled wave vector $R k_z$, for different partial wave indices $m=0,1,2$. Only two of the four eigenvalues are shown, as eigenvalues of $\mathbb{K}_{11}$ always appear in pairs $(\lambda,-\lambda)$. The permittivities are indicated in the plots. }
\end{figure}

A third validation example involves the eigenvalues of $\mathbb{K}_{11}$ and the scattering Green function $\mathbb{\Gamma}$ for a dielectric cylinder.  The latter is fully specified by the scattering T-operator $\mathbb{T}$ of the cylinder. It is known exactly and constitutes the only exact result for a curved dielectric body which couples electric and magnetic polarizations upon scattering \cite{Noruzifar:2012wk}. When $\mathbb{T}$ is known, one can use the relation \cite{bimonte2021} 
\be
\mathbb{\Gamma}({\bf r},{\bf r}')=\int \!d\tilde{\bf r}\int \!d\tilde{\bf r}' \, \mathbb{G}_0 ({\bf r},\tilde{\bf r}) \mathbb{T} (\tilde{\bf r},\tilde{\bf r}')\mathbb{G}_0 (\tilde{\bf r}',{\bf r}')
\ee 
to compute the components of $\mathbb{\Gamma}$ in a partial wave expansion of $\mathbb{G}_0$ where the integrations now extend over the volume of the cylinder. Specifically, vector cylindrical waves are a convenient choice to obtain the SSO $\mathbb{K}_{11}$ of the cylinder and to extract from the MSE for $\mathbb{\Gamma}$ the T-operator elements $\mathbb{T}^{\alpha \alpha'}(m,\kappa,k_z)$ for $\alpha,\alpha'\in \{E,H\}$, the imaginary wave number $\kappa=\xi/c$, the wave vector $k_z$ along the cylinder axis and  the angular quantum number $m$ (see  App.~D for details). The elements of $\mathbb{K}_{11}$ can be expressed in terms of Bessel functions $K_m(z)$, $I_m(z)$ but the expressions are too lengthy to be shown here. For each value $\kappa R$, $k_z R$ and integer partial wave index $m\ge 0$  there are four eigenvalues of $\mathbb{K}_{11}$. They can be also expressed in terms of Bessel functions. For all considered permittivities, frequencies and wave vectors, the moduli of the eigenvalues were found to be less than one. Fig.~\ref{fig:cylinder} shows the absolute values of the eigenvalues for different permittivities. For large $R\kappa$ the eigenvalues become independent of $Rk_z$ and $m$ as they approach the  high energy limit given by Eq.~(\ref{eighigh}).

\begin{figure}[h]
\includegraphics[width=.75\textwidth]{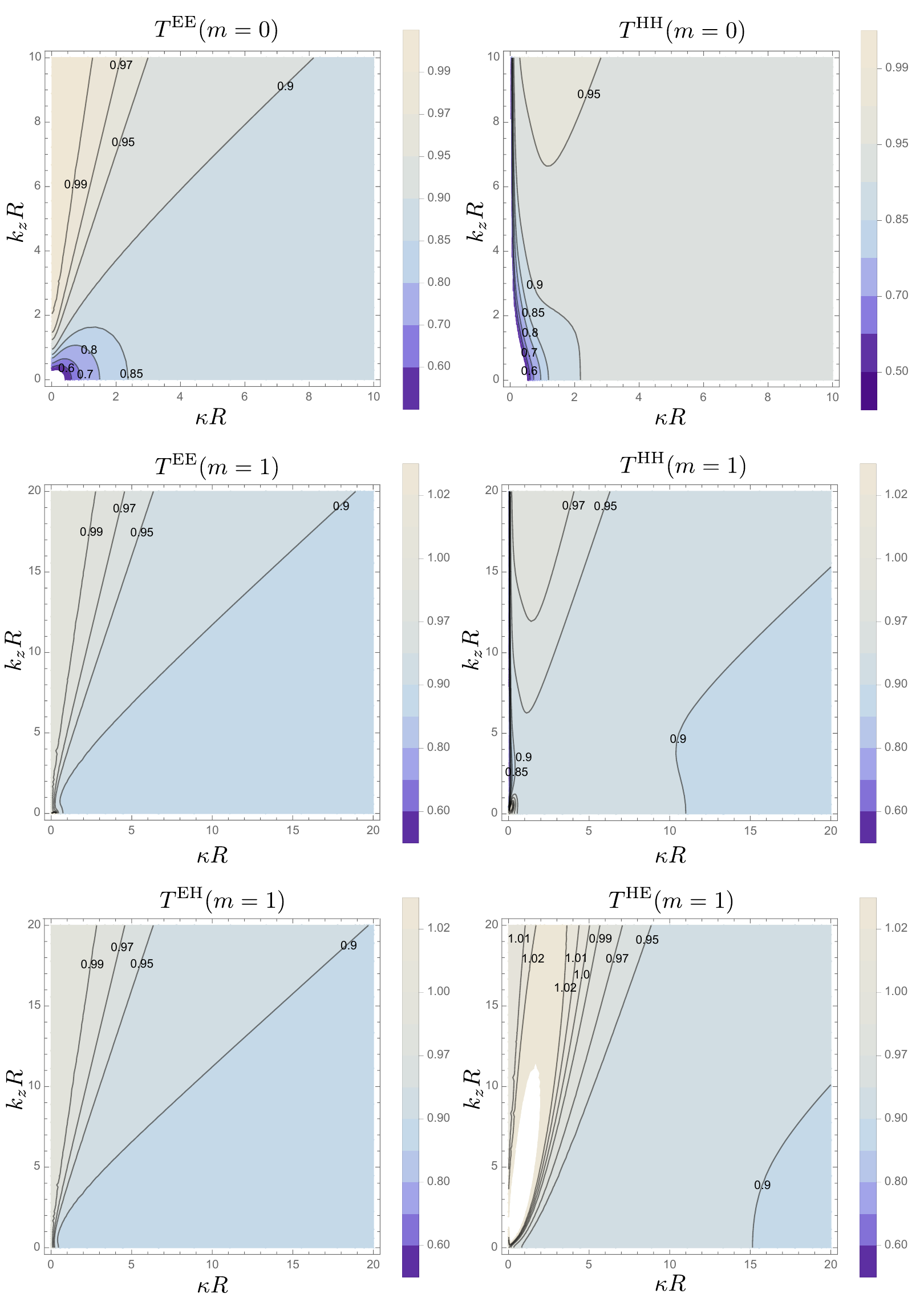} 
\caption{\label{fig:cylinder_gamma}
{\bf Multiple scattering expansion of the scattering Green function of a dielectric cylinder.}
Contour plots of the ratio of the T-matrix elements of a dielectric cylinder of radius $R$ computed with the MSE to order $p=3$ and the  exact results, as a function of re-scaled imaginary frequency $\kappa R$ and the re-scaled wave vector $k_z R$ along the cylinder axis. The dielectric permittivities of the cylinder are $\epsilon=30$ and $\mu=1$. Shown are the lowest order T-matrix elements with angular quantum numbers $m=0,1$ for all four combinations of polarizations E and H. (For $m=0$ the polarization couplings $T^{EH}$, $T^{HE}$ vanish.)
}
\end{figure}

Next, we study the scattering Green function. The panels in Fig.~\ref{fig:cylinder_gamma} display interesting aspects of the convergence of the approximant for $\mathbb{T}^{\alpha\alpha'}$ with the MSE $(1-\mathbb{K}_{11})^{-1}=\sum_{n=0}^p \mathbb{K}_{11}^p$ for order $p=3$. The contour plots show the ratio of the approximant and the exact T-operator elements for  $m=0,1$ as a function of the dimensionless wave numbers $\kappa R$, $k_z R$ for a cylinder of radius $R$ and permittivities $\epsilon_1=30$ and $\mu_1=1$.  While at this low order overall convergence has reached already agreement of better than $85\%$ with the exact result, the plots reveal a complex dependence of the convergence rate on wave numbers.  Typically convergence accelerates with decreasing frequency scale $\kappa$ and increasing wave number $k_z$, with the exception of lowest $m=0$ elements which show slow convergence around the static, long wave length limits $\kappa=k_z=0$. This slow down can be understood from the presence of a logarithmic divergence in $\mathbb{T}$ for $m=0$ which is a consequence of the infinite length of the cylinder \cite{Noruzifar:2012wk}. 
The observation of fast convergence of the MSE for $\mathbb{\Gamma}$ is important as it determines directly the Casimir-Polder interaction between a surface and a polarizable particle \cite{Casimir:1948jf}.

\section{Conclusion and discussion}

After decades of efforts  by many researchers, the power  of  integral equations methods \cite{chew,volakis}    in computational electromagnetism  is by now an established fact.   Only recently, however,  these methods have been  these applied to Casimir physics  \cite{reid2013,rodriguez}. The findings of  \cite{reid2013,rodriguez} undoubtedly represent a significant  progress in the field, because they make possible for the first time to compute, at least in principle, Casimir interactions  for  arbitrary arrangements of any number of (homogeneous) magneto-dielectric bodies of any  shape. This is very important in view of applications to micro and nano mechanical devices of complex shapes, where the Casimir force may play an important role.
While it is a huge step forward, the approach of  \cite{reid2013,rodriguez}    suffers from the drawback that its implementation  is extremely costly in terms of the required computer resources, which may not be generally available to the interested researchers.

In \cite{emig2023}  we introduced a whole new class of exact   integral-equation representations of the Casimir and Casimir-Polder  interaction between bodies of arbitrary shape and material composition. The present work offers a detailed and pedagogical presentation of our methods, which may not be familiar to the majority of researchers in Casimir physics. A major  difference with respect to \cite{reid2013,rodriguez} is that in our approach the Casimir and  Casimir-Polder  interactions are expressed in term of  surface integral equations of the 2nd Fredholm type,  which  are amenable to a  MSE in terms of elementary free-space propagators. We underline that  our  representation does not depend on the scattering amplitude of the bodies. Moreover, our semi-analytical MSE does not involve a discrete mesh representation of the geometry and requires no numerical computation and inversion of large matrices over boundary elements, a computationally expensive task.  In our approach,  the interaction is in fact expressed in terms of iterated integrals of elementary functions extended on the surfaces of the bodies.  For soft material bodies like those usually considered in biological systems, the  MSE converges quickly, and then already the first terms of the expansion may provide  a fairly accurate estimate of the interaction energy. The Si-Au wedge-plate system studied in \cite{emig2023} shows that even in the case of condensed bodies convergence of the  MSE is rather fast. We believe that the possibility of getting, via the  MSE,   an estimate of the Casimir energy in a complex geometry, by just performing  simple surface integrals,  adds a useful tool to the toolbox  of researchers in the field.

We  envisage several possible future directions for our work. One important advantage of our approach is that our representations of the Casimir and Casimir-Polder  interactions involve several free parameters, that may be  in principle adjusted to the dielectric properties of the bodies, in order to speed convergence of the expansion.  The problem of determining the optimal choice of these coefficients is a very interesting topic, that we plan to investigate in future publications.
Another clear direction is to apply the MSE in the real frequency domain, to the technologically important problem of radiative transfer at the micro and nano scales, a subject of intense study in recent years \cite{volokitin,biehs2021}. Our  approach can be easily adapted to this problem, by following steps similar to those of \cite{rodriguez2013}.  The  non-trivial issue that requires a systematic investigation is the domain of convergence of  the MSE  for real-frequencies, for the materials and the frequency ranges that are relevant to the problem.

In conclusion, our rapidly convergent MSE can provide a powerful tool to delve deeper into Casimir and thermal phenomena in sub-micrometre structures composed of various materials which cannot be understood by simple additive power laws and planar or spherical surface interactions.

\newpage 

\appendix

\section{Surface-integral formulation of electromagnetic scattering}

In this Appendix, we briefly review  surface-integral formulation of  EM scattering by dielectric objects. This  provides a convenient basis for the derivation of the SSO, which is the subject of the next Appendix. 

We start from the formulation of our scattering problem. Let us consider a collection of $N$ dielectric bodies, characterized by the respective (frequency-dependent) permittivities $\epsilon_{\sigma}, \mu_{\sigma}$, embedded in a dielectric medium with permittivities $\epsilon_{0}, \mu_{0}$. We let $V_{\sigma}$ the volume occupied by  the $\sigma$-th body, and $S_{\sigma}$ its surface, with  $\hat{\bf n}_{\sigma}({\bf u})$ the unit outward normal  to  $S_{\sigma}$.  We finally denote by $V_0$ the region of space, outside the collection of $N$ bodies.   We imagine a  distribution of electric and magnetic sources $({\bf J},{\bf M})$   in $V_0$,   and we let $({\bf E}_{\rm inc},{\bf H}_{\rm inc})$  the incident EM field  radiated (in the absence of the N bodies) by $({\bf J},{\bf M})$:
\be 
( {\bf E}_{\rm inc} , {\bf H}_{\rm inc}) ( {\bf r}) =  \int_{V_0} d {\bf r}'\,  \mathbb{G}_0 ({\bf r}-{\bf r}')\cdot ( {\bf J} , {\bf M}) ({\bf r}')  \;,
\ee
where $ \mathbb{G}_{0}$  denote the Green tensors   for a homogeneous and isotropic medium with  permittivities $\epsilon_0$, $\mu_0$, respectively (the explicit expression of the Green tensors are provided in Appendix \ref{sec:green}).
Solution of the N-body scattering problem  requires solving Maxwell Equations in the regions $V_0,V_1,\cdots V_N$,  with sources $({\bf J},{\bf M})$ in $V_0$,    subjected to the boundary  conditions  that the tangential components of  the EM field ${\bf E}$ and ${\bf H}$ are continuous across the $N$ surfaces $S_{\sigma}$:
\begin{eqnarray}
\hat{\bf n}_{\sigma}({\bf u}) \times {\bf E}_+({\bf u}) &=&\hat{\bf n}_{\sigma}({\bf u}) \times {\bf E}_- ({\bf u})\,, \nonumber \\
\hat{\bf n}_{\sigma}({\bf u}) \times {\bf H}_+({\bf u})&=&\hat{\bf n}_{\sigma}({\bf u}) \times {\bf H}_-({\bf u})\;,\label{bc}
\end{eqnarray}
where   ${\bf E}_+$  and ${\bf E}_-$  (${\bf H}_+$  and ${\bf H}_-$) denote, respectively,  the values of the electric (magnetic) field at points just outside and inside the surface $S_{\sigma}$. It is convenient to define the electric and magnetic "surface currents'' $ {\bf j}_{\sigma}({\bf u})$ and $ {\bf m}_{\sigma}({\bf u})$, with ${\bf u} \in S_{\sigma}$, by the relations:
\begin{eqnarray} 
 {\bf j}_{\sigma}({\bf u})   &\equiv&  \hat{\bf n}_{\sigma}({\bf u}) \times {\bf H}({\bf u})\;,\nonumber\\
 {\bf m}_{\sigma}({\bf u})  &\equiv&- \hat{\bf n}_{\sigma}({\bf u}) \times {\bf E}({\bf u})\;,
\end{eqnarray} 
By using Green's theorem \cite{born,maradudin,Harrington:2001yb}, one can prove the following four sets of integral identities,  which  relate  the EM field  ${\bf E}$ and ${\bf H}$  to the incident field  $({\bf E}_{\rm inc},{\bf H}_{\rm inc})$ and to the boundary fields  ${\bf j}_{1},\dots, {\bf m}_{N}$:
\begin{eqnarray}
 ({\bf E}, {\bf H}) ({\bf r}) &=& ( {\bf E}_{\rm inc} ,  {\bf H}_{\rm inc}) ({\bf r})+ ( {\tilde {\bf E}}^{(0)} , {\tilde {\bf H}}^{(0)} ) ( {\bf r}; \;  {\bf j}_{1},\dots, {\bf m}_{N})  \;,\;\;\;{\bf r} \in V_0\;,\label{scatext} \\ 
0&=&  ( {\bf E}_{\rm inc} ,  {\bf H}_{\rm inc})( {\bf r} ) + ( {\tilde {\bf E}}^{(0)} , {\tilde {\bf H}}^{(0)} ) ( {\bf r};\; {\bf j}_{1},\dots,  {\bf m}_{N})\;,  \;\;\;{\bf r} \notin V_0\;,
   \label{identit0}
\end{eqnarray}
\begin{eqnarray}
 ({\bf E}, {\bf H}) ({\bf r}) &=&-  ( {\tilde {\bf E}}^{(\sigma)} , {\tilde {\bf H}}^{(\sigma)} )  ({\bf r}; \; {\bf j}_{\sigma},{\bf m}_{\sigma})\;,\;\;\;{\bf r} \in V_{\sigma} \label{scatint}\\
0&=&  \;\;\; ( {\tilde {\bf E}}^{(\sigma)} , {\tilde {\bf H}}^{(\sigma)} ) ( {\bf r};\;{\bf j}_{\sigma},{\bf m}_{\sigma})\;, \;\;\;{\bf r} \notin V_{\sigma}
.\label{identit1}
\end{eqnarray}
In the above relations,  $ {\tilde {\bf E}}^{(\rho)} , {\tilde {\bf H}}^{(\rho)}$ ($\rho=0,1,\cdots,N$) denote the  following surface integrals:
 \begin{eqnarray}
( {\tilde {\bf E}}^{(0)} , {\tilde {\bf H}}^{(0)} )( {\bf r}; \; {\bf j}_{1},\dots, {\bf m}_{N})& \equiv & \sum_{\sigma=1}^N \int_{S_\sigma} d s_{{\bf u}}\, \mathbb{G}_0 ({\bf r}-{\bf u}) \cdot ( {\bf j}_{\sigma} ,{\bf m}_{\sigma} )({\bf u})  \;,\nonumber\\
( {\tilde {\bf E}}^{(\sigma)} ,  {\tilde {\bf H}}^{(\sigma)} )( {\bf r}; \;{\bf j}_{\sigma},{\bf m}_{\sigma})& \equiv & \int_{S_\sigma}  d s_{{\bf u}} \, \mathbb{G}_{\sigma} ({\bf r}-{\bf u})\cdot  ( {\bf j}_{\sigma}, {\bf m}_{\sigma})({\bf u}) \;, \label{defEsigmabis}
\end{eqnarray}
where $ d s_{{\bf u}}$ is the area element on $S_\sigma$, while $ \mathbb{G}_{\sigma}^{(\alpha \beta)}$, $\sigma=0,1,\cdots,N$  denote the Green tensors   for a homogeneous and isotropic medium with frequency dependent electric and magnetic permittivities $\epsilon_\sigma(\omega)$, $\mu_\sigma(\omega)$, respectively.

Independent of the Green's theorem, validity of the identities Eq.(\ref{scatext} - \ref{identit1}) can be easily understood by using a nice mathematical trick, that goes by the name of the "equivalence principle'' \cite{Harrington:2001yb}. The trick consists in  introducing  the following  $N+1$  EM fields $({\bf E}^{(0)},{\bf H}^{(0)}), \cdots,({\bf E}^{(N)},{\bf H}^{(N)})$: 
\be
({\bf E}^{(0)},{\bf H}^{(0)}) ({\bf r})  = \left\{ \begin{array}{ll}
({\bf E},{\bf H}) ({\bf r})	\;,  &	\;	\; {\bf r} \in V_0\\
 ( {\bf 0}, {\bf 0}) & \;	\; {\bf r} \notin V_0\\
\end{array}\right. \, .\label{Ezero}
\ee  
\be
({\bf E}^{(\sigma)},{\bf H}^{(\sigma)}) ({\bf r})  = \left\{ \begin{array}{ll}
({\bf E},{\bf H}) ({\bf r})	\;,  &	\;	\; {\bf r} \in V_{\sigma}\\
 ( {\bf 0}, {\bf 0}) & \;	\; {\bf r} \notin V_{\sigma}\\
\end{array}\right. \, .\label{Esigmater}
\ee  
As we see, the field $({\bf E}^{(0)},{\bf H}^{(0)})$
coincides with the actual EM field $({\bf E},{\bf H})$ at points  in the medium surrounding the bodies, and it {\it vanishes} at all points {\it inside} bodies. 
Vice-versa,  each of the fields $({\bf E}^{(\sigma)},{\bf H}^{(\sigma)})$   coincides with the  total field $({\bf E},{\bf H})$ at points {\it inside} the  respective body, and {\it vanishes} at all other points  of space. All these fields are clearly {\it unphysical}, since they  do not fulfill the  boundary conditions Eq. (\ref{bc})  on at least one among the surfaces $S_{\sigma}$.  While unphysical, these fields have by construction  the nice property of being solutions of Maxwell Equations in infinite {\it homogeneous} space, with {\it constant} dielectric properties.   More precisely, the field $({\bf E}^{(0)},{\bf H}^{(0)})$ satisfies   (except on the surfaces $(S_{1}, \cdots S_N)$, where it is discontinuous) Maxwell Equations in a 
medium having {\it everywhere} the  permittivities $(\epsilon_0,\mu_0)$ of the medium surrounding the bodies, while each of   the  fields $({\bf E}^{(\sigma)},{\bf H}^{(\sigma)})$ satisfies   (except on the surface of the $\sigma$-th body, where it is discontinuous) Maxwell Equations in a 
medium having {\it everywhere} the  permittivities $(\epsilon_{\sigma},\mu_{\sigma})$ of the material filling the $\sigma$-th body.  Now comes the main observation. Since the media in which all these fields live are spatially homogeneous,   one concludes that these fields   are in fact free fields, and therefore  they can be expressed as convolutions of  free-space Green tensors with the appropriate sources.   By construction,  the  sources  of  $({\bf E}^{(0)},{\bf H}^{(0)})$ are the original external sources $({\bf J}, {\bf M})$ of our scattering problem, together with the  2N surface currents $({\bf j}_1,\cdots {\bf m}_N)$ arising from the discontinuity of $({\bf E}^{(0)},{\bf H}^{(0)})$ across the bodies surfaces.  The identities  in Eq. (\ref{scatext}) and (\ref{identit0}) become obvious, if one realizes that they represent the expression of $({\bf E}^{(0)},{\bf H}^{(0)})$ as a convolution of $\mathbb{G}_0$ with its sources  $({\bf J}, {\bf M})$  and $({\bf j}_1,\cdots {\bf m}_N)$.  An analogous argument applies to the  fields $({\bf E}^{(\sigma)},{\bf H}^{(\sigma)})$.   From the discontinuity of $({\bf E}^{(\sigma)},{\bf H}^{(\sigma)})$ across $S_{\sigma}$, one sees that  $({\bf E}^{(\sigma)},{\bf H}^{(\sigma)})$  is sourced by the surface currents $(-{\bf j}_{\sigma}, -{\bf m}_{\sigma})$.  Upon expressing   $({\bf E}^{(\sigma)},{\bf H}^{(\sigma)})$ as a convolution of  $\mathbb{G}_{\sigma}$ with $(-{\bf j}_{\sigma}, -{\bf m}_{\sigma})$,   one recovers at once the identities  in Eq. (\ref{scatint}) and (\ref{identit1}). 

Let us go back now to Eq. (\ref{scatext}):  this integral relation  shows that at points ${\bf r}$ outside the bodies, the scattered field  $(  {\bf E}_{\rm scat},   {\bf H}_{\rm scat} )$ 
coincides with the surface integral $ ( {\tilde {\bf E}}^{(0)} , {\tilde {\bf H}}^{(0)} ) $
\be
(  {\bf E}_{\rm scat},   {\bf H}_{\rm scat} ) ( {\bf r}) = ( {\tilde {\bf E}}^{(0)} , {\tilde {\bf H}}^{(0)} ) ( {\bf r}; \;  {\bf j}_{1},\dots, {\bf m}_{N})  \;,\;\;\;{\bf r} \in V_0\;.\label{scatext2} 
\ee
This relation shows that the scattering problem is solved, provided that the 2N surface currents  $({\bf j}_{1},\dots, {\bf m}_{N})$ can be actually computed. 
In the next Appendix, we show  how this goal can be achieved, using the SSO.

\section{Alternative derivation of the SSO}
\label{sec:SSO}

In this Appendix, we  construct the SSO that allows to compute the surface currents providing the solution of the EM scattering problem.    The starting point is provided by the identities in Eqs. (\ref{identit0}) and (\ref{identit1}). 
Upon taking the limits of Eqs.(\ref{identit0}) and (\ref{identit1}) as the point ${\bf r}$ approaches the point ${\bf u}$ on the surface $S_{\sigma}$, and then taking a vector product with the unit normal to  $S_{\sigma}$, one obtains the following identities:
\begin{eqnarray}
\hat{\bf n}_{\sigma}({\bf u}) \times {\tilde {\bf E}}^{(0)}_-( {\bf u};\; {\bf j}_{1},\dots,  {\bf m}_{N})&+&\hat{\bf n}_{\sigma} \times {\bf E}_{\rm inc}( {\bf u} )=0\;,\nonumber\\
\hat{\bf n}_{\sigma}({\bf r}) \times {\tilde {\bf H}}^{(0)}_-( {\bf u};\; {\bf j}_{1},\dots, {\bf m}_{N})&+& \hat{\bf n}_{\sigma} \times {\bf H}_{\rm inc}({\bf u})=0\;,\nonumber\\
\hat{\bf n}_{\sigma}({\bf u}) \times {\tilde {\bf E}}^{(\sigma)}_+( {\bf u};\;{\bf j}_{\sigma},{\bf m}_{\sigma})&=&0\;,\nonumber\\
\hat{\bf n}_{\sigma}({\bf u}) \times {\tilde {\bf H}}^{(\sigma)}_+( {\bf u};\;{\bf j}_{ \sigma},{\bf m}_{\sigma})&=&0\;.\label{identit}
\end{eqnarray}
The above relations constitute an overdetermined set of $4 N$ integral Equations in the $2 N$ unknown boundary fields
$( {\bf j}_{1},\dots, {\bf m}_{N})$.   A consistent set of Equations can be obtained  by taking $2N$ distinct linear combinations of the $4 N$ Equations (\ref{identit}):
\begin{eqnarray}
C^{(e|E)}_{\sigma} \hat{\bf n}_{\sigma} \times {\tilde {\bf E}}^{(0)}_- ( {\bf j}_{1},\dots, {\bf m}_{N})- C^{(i|E)}_{\sigma} \hat{\bf n}_{\sigma} \times {\tilde {\bf E}}^{(\sigma)}_+({\bf j}_{\sigma},{\bf m}_{\sigma})&=&- C^{(e|E)}_{\sigma} \hat{\bf n}_{\sigma} \times { {\bf E}}_{\rm inc}\;,\nonumber\\
C^{(e|H)}_{\sigma} \hat{\bf n}_{\sigma} \times {\tilde {\bf H}}^{(0)}_-( {\bf j}_{1},\dots, {\bf m}_{N})- C^{(i|H)}_{\sigma} \hat{\bf n}_{\sigma} \times {\tilde {\bf H}}^{(\sigma)}_+({\bf j}_{\sigma},{\bf m}_{\sigma})&=&- C^{(e|H)}_{\sigma} \hat{\bf n}_{\sigma} \times { {\bf H}}_{\rm inc}\;,\label{inteq0}
\end{eqnarray} 
where for brevity we do not display the explicit dependence of the boundary fields on the point ${\bf u}$. We remark that the coefficients in Eq. (\ref{inteq0})  are defined up to rescalings by arbitrary non-vanishing factors $\lambda^{(\alpha)}_{\sigma}$:
\begin{eqnarray}
(C^{(i|\alpha)}_{\sigma},C^{(e|\alpha)}_{\sigma})& \rightarrow& \lambda_{\sigma}^{(\alpha)}\,(C^{(i|\alpha)}_{\sigma},C^{(e|\alpha)}_{\sigma})\;.
\end{eqnarray}
It is  convenient to re-express Eqs.(\ref{inteq0})  in terms of  the values of the surface integrals $( {\tilde {\bf E}}^{(\rho)},  {\tilde {\bf H}}^{(\rho)})$ computed directly on the surfaces $S_{\sigma}$.  This can be done by observing that, for an arbitrary choice of the surface currents,  the surface integrals $( {\tilde {\bf E}}^{(0)}, {\tilde {\bf H}}^{(0)})$,  $( {\tilde {\bf E}}^{(\sigma)}, {\tilde {\bf H}}^{(\sigma)})$ satisfy the jump conditions:
\begin{eqnarray}
\hat{\bf n}_{\sigma}({\bf u}) \times \left[{\tilde  {\bf E}}^{(0)}_+({\bf u})- {\tilde {\bf E}}^{(0)}_- ({\bf u}) \right] &=& -{\bf m}_{\sigma} ({\bf u})\,, \nonumber \\
\hat{\bf n}_{\sigma}({\bf u}) \times \left[{\tilde  {\bf E}}^{(\sigma)}_+({\bf u})- {\tilde {\bf E}}^{(\sigma)}_- ({\bf u}) \right] &=& -{\bf m}_{\sigma} ({\bf u})\,, \nonumber \\
\hat{\bf n}_{\sigma}({\bf u}) \times \left[ {\tilde {\bf H}}^{(0)}_+({\bf u})- {\tilde {\bf H}}^{(0)}_- ({\bf u}) \right] &=& {\bf j}_{\sigma} ({\bf u})\,, \nonumber \\
\hat{\bf n}_{\sigma}({\bf u}) \times \left[ {\tilde {\bf H}}^{(\sigma)}_+({\bf u})- {\tilde {\bf H}}^{(\sigma)}_- ({\bf u}) \right] &=& {\bf j}_{\sigma} ({\bf u})
 \;,\label{jump}
\end{eqnarray}
On the other hand, one know  \cite{maradudin,muller} that the fields   $({\tilde {\bf E}}^{(\rho)},{\tilde {\bf H}}^{(\rho)})({\bf u})$,    are the averages of the corresponding values just inside and outside $S_{\sigma}$:
\begin{eqnarray}
\hat{\bf n}_{\sigma}({\bf u}) \times \left[ {\tilde {\bf E}}^{(0)}_+({\bf u})+ {\tilde {\bf E}}^{(0)}_- ({\bf u}) \right] &=&  2\; \hat{\bf n}_{\sigma}({\bf u}) \times {\tilde {\bf E}}^{(0)}({\bf u}) \,, \nonumber \\
\hat{\bf n}_{\sigma}({\bf u}) \times \left[ {\tilde {\bf E}}^{(\sigma)}_+({\bf u})+ {\tilde {\bf E}}^{(\sigma)}_- ({\bf u}) \right] &=& 2\; \hat{\bf n}_{\sigma}({\bf u}) \times  {\tilde{\bf E}}^{(\sigma)}({\bf u}) \,, \nonumber \\
\hat{\bf n}_{\sigma}({\bf u}) \times \left[ {\tilde {\bf H}}^{(0)}_+({\bf u})+{\tilde  {\bf H}}^{(0)}_- ({\bf u}) \right] &=& 2\; \hat{\bf n}_{\sigma}({\bf u}) \times {\tilde {\bf H}}^{(0)}({\bf u}) \,, \nonumber \\
\hat{\bf n}_{\sigma}({\bf u}) \times \left[ {\tilde {\bf H}}^{(\sigma)}_+({\bf u})+{\tilde  {\bf H}}^{(\sigma)}_- ({\bf u}) \right] &=& 2\; \hat{\bf n}_{\sigma}({\bf u}) \times  {\tilde {\bf H}}^{(\sigma)}({\bf u}) 
 \;.\label{avefi} 
\end{eqnarray}
The above Equations can be used to eliminate  ${\tilde {\bf E}}^{(0)}_-,{\tilde{\bf H}}^{(0)}_-, {\tilde {\bf E}}^{(\sigma)}_+, {\tilde {\bf H}}^{(\sigma)}_+$ from Eqs. (\ref{inteq0}). By doing so, one arrives at the following set of integral equations for the surface currents:
\be
\!\!\!\!\!\!\!({C^{(e|H)}_{\sigma}+C^{(i|H)}_{\sigma}})\,{\bf j}_{\sigma} - 2\, {C^{(e|H)}_{\sigma}} \hat{\bf n}_{\sigma} \times {\tilde {\bf H}}^{(0)}( {\bf j}_{1},\dots, {\bf m}_{N}) + 2\,
{C^{(i|H)}_{\sigma}}   \hat{\bf n}_{\sigma} \times {\tilde {\bf H}}^{(\sigma)}({\bf j}_{\sigma},{\bf m}_{\sigma}) =
 2\, {C^{(e|H)}_{\sigma}}  \hat{\bf n}_{\sigma} \times {\bf H}_{\rm inc}\;,\nonumber
 \ee
 \be
\!\!\!\!\!\!\!({C^{(e|E)}_{\sigma}+C^{(i|E)}_{\sigma}})\,{\bf m}_{\sigma} + 2\, {C^{(e|E)}_{\sigma}}  \hat{\bf n}_{\sigma} \times {\tilde {\bf E}}^{(0)}( {\bf j}_{1},\dots, {\bf m}_{N}) - 2\,
{C^{(i|E)}_{\sigma}}  \hat{\bf n}_{\sigma} \times {\tilde {\bf E}}^{(\sigma)} ({\bf j}_{\sigma},{\bf m}_{\sigma})= - 2\, {C^{(e|E)}_{\sigma}}   \hat{\bf n}_{\sigma} \times {\bf E}_{\rm inc}\;. 
\label{inteq1}
\ee
For generic values of the coefficients, both ${C^{(e|H)}_{\sigma}+C^{(i|H)}_{\sigma}}$ and ${C^{(e|E)}_{\sigma}+C^{(i|E)}_{\sigma}}$ are different from zero, and then
the integral Equations (\ref{inteq1}) can be recast in the form of Eq. (\ref{eq:Fredholm_currents}).
The proof that   Eqs. (\ref{inteq1}) actually determine uniquely the surface currents $({\bf j}_{1},\dots, {\bf m}_{N})$ at all  complex frequencies, and for any choice of the $4 N$ coefficients  $(C^{(e|E)}_{\sigma}, C^{(e|H)}_{\sigma}, C^{(i|E)}_{\sigma}, C^{(i|H)}_{\sigma})$, such that both ${C^{(e|H)}_{\sigma}+C^{(i|H)}_{\sigma}}$ and ${C^{(e|E)}_{\sigma}+C^{(i|E)}_{\sigma}}$ are different from zero,  can indeed be  obtained by a simple adaptation  of  the proof given in \cite{muller} for a single body and   for the  particular choice of coefficients, denoted by (C1) in Sec.~IV. 



\section{Perfect conductors}
\label{sec:PM}

In this Appendix we  work out the SSO for a collection of perfect conductors. The scattering problem now involves a system of $N$ perfectly conducting bodies placed in a medium characterized by electric and magnetic permittivities $\epsilon_{0}, \mu_{0}$, respectively. Like before,  we imagine a  distribution of electric and magnetic sources $({\bf J},{\bf M})$   in the region $V_0$ outside the conductors. Solution of the N-body scattering problem  now requires solving Maxwell Equations in the region $V_0$,  with sources $({\bf J},{\bf M})$ in $V_0$,    subjected to the boundary  conditions  that the tangential component of  the electric field ${\bf E}$ vanishes on the boundaries of the conductors:
\be
\hat{\bf n}_{\sigma}({\bf u}) \times {\bf E}_+({\bf u}) =0\;.\label{PMbc}
\ee
In view of this simple condition, we now have only one set of surface currents, namely the   electric currents 
\begin{equation} 
 {\bf j}_{\sigma}({\bf u})   =  \hat{\bf n}_{\sigma}({\bf u}) \times {\bf H}_+({\bf u})\; .
  \end{equation} 
By Green's theorem \cite{born,maradudin,Harrington:2001yb}, one  finds  the following two sets of integral identities,  which  relate  the EM field  ${\bf E}$ and ${\bf H}$  to the external field  $({\bf E}_{\rm inc},{\bf H}_{\rm inc})$ and to the boundary fields  ${\bf j}_{1},\dots, {\bf j}_{N}$:
\begin{eqnarray}
 ({\bf E}, {\bf H}) ({\bf r}) &=& ( {\bf E}_{\rm inc} ,  {\bf H}_{\rm inc}) ({\bf r})+ ( {\tilde {\bf E}}^{(0)} , {\tilde {\bf H}}^{(0)} ) ( {\bf r}; \;  {\bf j}_{1},\dots, {\bf j}_{N})  \;,\;\;\;{\bf r} \in V_0\;,\label{scatextPM} \\ 
0&=&  ( {\bf E}_{\rm inc} ,  {\bf H}_{\rm inc})( {\bf r} ) + ( {\tilde {\bf E}}^{(0)} , {\tilde {\bf H}}^{(0)} ) ( {\bf r};\; {\bf j}_{1},\dots,  {\bf j}_{N})\;,  \;\;\;{\bf r} \notin V_0\;,
   \label{identit0PM}
\end{eqnarray}
The above Equations show that the PM scattering problem is solved if one can determine the $N$ surface currents ${\bf j}_{1},\dots, {\bf j}_{N}$. Proceeding as  in the case of dielectric bodies, we consider the limits of Eq. (\ref{identit0PM}) as ${\bf r}$ tends to a point ${\bf u}$ on the surfaces of the conductors. This gives us:
\begin{eqnarray}
\hat{\bf n}_{\sigma}({\bf u}) \times {\tilde {\bf E}}^{(0)}_-( {\bf u};\; {\bf j}_{1},\dots,  {\bf j}_{N})&+&\hat{\bf n}_{\sigma} \times {\bf E}_{\rm inc}( {\bf u} )=0\;,\nonumber\\
\hat{\bf n}_{\sigma}({\bf r}) \times {\tilde {\bf H}}^{(0)}_-( {\bf u};\; {\bf j}_{1},\dots, {\bf j}_{N})&+& \hat{\bf n}_{\sigma} \times {\bf H}_{\rm inc}({\bf u})=0\;.\label{identitPM}
\end{eqnarray}
The above relations constitute an overdetermined set of $2 N$ integral Equations in the $N$ unknown boundary fields
$( {\bf j}_{1},\dots, {\bf j}_{N})$.   A consistent set of Equations can be obtained  by taking $N$ distinct linear combinations of the $2 N$ Equations (\ref{identitPM}):
\begin{eqnarray}
&&C^{(e|E)}_{\sigma} \hat{\bf n}_{\sigma} \times {\tilde {\bf E}}^{(0)}_-( {\bf j}_{1},\dots, {\bf j}_{N}) + C^{(e|H)}_{\sigma} \hat{\bf n}_{\sigma} \times {\tilde {\bf H}}^{(0)}_- ( {\bf j}_{1},\dots, {\bf j}_{N}) \nonumber \\
&&\;\;\;\;\;=- C^{(e|E)}_{\sigma} \hat{\bf n}_{\sigma} \times { {\bf E}}_{\rm inc} -C^{(e|H)}_{\sigma} \hat{\bf n}_{\sigma} \times { {\bf H}}_{\rm inc}\;,\label{inteq0PM}
\end{eqnarray} 
Similar to what we did earlier, we can take advantage of the identities in the last two lines of   Eqs. (\ref{jump}) and (\ref{avefi}) to re-express the above integral Equation in terms of the values of ${\tilde {\bf E}}^{(0)}$ and ${\tilde {\bf H}}^{(0)}$ on the surfaces $S_{\sigma}$:
\begin{eqnarray}
&&C^{(e|H)}_{\sigma}  \,{\bf j}_{\sigma} - 2\, {C^{(e|H)}_{\sigma}} \hat{\bf n}_{\sigma} \times {\tilde {\bf H}}^{(0)}( {\bf j}_{1},\dots, {\bf j}_{N}) +
2\, {C^{(e|E)}_{\sigma}}  \hat{\bf n}_{\sigma} \times {\tilde {\bf E}}^{(0)}( {\bf j}_{1},\dots, {\bf j}_{N})  \nonumber\\ 
&&\;\;\;\;\;\;=
 2\, {C^{(e|H)}_{\sigma}}  \hat{\bf n}_{\sigma} \times {\bf H}_{\rm inc}  - 2\, {C^{(e|E)}_{\sigma}}   \hat{\bf n}_{\sigma} \times {\bf E}_{\rm inc}\;. 
\label{inteq1PM}
 \end{eqnarray}
As in the general case for magneto-dielectric bodies, several different formulations  exist for the perfectly conducting limit, depending on the choice of the coefficients in Eq.~(\ref{inteq1PM}).  A possible  choice   is
\be
C^{(e|H)}_{\sigma} =0\;,\;\;\;\;C^{(e|E)}_{\sigma}=1\;.
\ee
The resulting integral equation reads
\be
  \hat{\bf n}_{\sigma} \times {\tilde {\bf E}}^{(0)}( {\bf j}_{1},\dots, {\bf j}_{N})  =
   -    \hat{\bf n}_{\sigma} \times {\bf E}_{\rm inc}\;,
\ee
Upon taking the vector product with $  \hat{\bf n}_{\sigma}$ of both members of the above Equation, we obtain the following integral equation for perfect conductors
\begin{equation}
\sum_{\sigma'=1}^N\int_{S_{\sigma'}}\!\!ds_{{\bf u}'} \, \mathbb{B}^{(\rm PC)} _{\sigma\sigma'}({\bf u},{\bf u}')\;  {\bf j}_{\sigma'}  ({\bf u}')
= \!\int d{\bf r} \,\tilde{\mathbb{M}}^{(\rm PC)} _{\sigma}({\bf u},{\bf r})   {\bf J} (\bf{r}) \;,\label{johnPC}
\end{equation}
where
\begin{equation}
\begin{aligned}
\mathbb{B}^{(\rm PC)}_{\sigma\sigma'}({\bf u},{\bf u}') & =   \left[\mathbb{G}_0^{(EE)}({\bf u},{\bf u}') \right]_t\\
 \end{aligned} \;,\label{PMPMC}
\end{equation} 
\begin{equation}
\begin{aligned}
\tilde{\mathbb{M}}^{(\rm PC)}_{\sigma}({\bf u},{\bf r}) & = -  \left[\mathbb{G}_0^{(EE)}({\bf u},{\bf r})  \right]_t\\
 \end{aligned} \;.
\end{equation}
The integral equation is not of Fredholm form, and therefore it does not allow for a MSE. We note that this formulation was   used in a numerical investigation of the Casimir effect in \cite{reid2013}.  
We now consider the alternative choice
\be
C^{(e|H)}_{\sigma} =1\;,\;\;\;\;C^{(e|E)}_{\sigma}=0\;,
\ee 
which leads to the following integral equation of 2nd Fredholm type:
\begin{equation}
\label{eq:PC3}
\sum_{\sigma'=1}^N\int_{S_{\sigma'}}\!\!ds_{{\bf u}'} \,\left[\mathbb{1} - \mathbb{K}^{(\rm PC)}_{\sigma\sigma'}({\bf u},{\bf u}')\right]  {\bf j}_{\sigma'}  ({\bf u}')
= \!\int d{\bf r} \,\mathbb{M}^{(\rm PC)}_{\sigma}({\bf u},{\bf r})  {\bf J}  ({\bf r})
\end{equation}
 with
 \be
\mathbb{K}^{(\rm PC)}_{\sigma\sigma'}({\bf u},{\bf u}')  =  2 \,{\bf n}_\sigma({\bf u}) \times  \mathbb{G}_0^{(HE)}({\bf u},{\bf u}') \;, \label{PMK}
\ee
and
\be
\mathbb{M}^{(\rm PC)}_{\sigma}({\bf u},{\bf r})  =  2 \,{\bf n}_\sigma({\bf u}) \times  \mathbb{G}_0^{(HE)}({\bf u},{\bf r}) \;. 
\ee
These is the integral equation for PC used in \cite{balian1977,balian1978}.




\section{T-matrix of a magneto-dielectric cylinder}
\label{sec:cylin}

The T-operator of a dielectric cylinder of radius $R$ assumes a $2\times 2$ block diagonal form in vector cylindrical waves labelled by the angular quantum number $m$ and the wave vector $k_z$ along the cylinder axis \cite{rahi2009}. It is assumed that the cylinder has electric and magnetic permittivities $\epsilon$ and $\mu$, and the surrounding medium is vacuum ($\epsilon_0=\mu_0=1$).
On the imaginary frequency axis, and with $p_0=\sqrt{\kappa^2+k_z^2}$, $p_1=\sqrt{\epsilon\mu\kappa^2+k_z^2}$, the diagonal elements are given by 
\cite{Noruzifar:2012wk}
\begin{equation}
\begin{aligned}
  \mathbb{T}^{HH}(m,\kappa,k_z) &= -\frac{I_m(p_0 R)}{K_m(p_0 R)} \frac{\Delta_1\Delta_4 +K^2}{\Delta_1\Delta_2+K^2}\,,\\
  \mathbb{T}^{EE}(m,\kappa,k_z) &= -\frac{I_m(p_0 R)}{K_m(p_0 R)} \frac{\Delta_2\Delta_3 +K^2}{\Delta_1\Delta_2+K^2}\,,\\
  \mathbb{T}^{HE}(m,\kappa,k_z) &= -\mathbb{T}^{EH}(m,\kappa,k_z)=  \frac{K}{ \sqrt{\epsilon\mu} (p_0R)^2 K_m(p_0 R)^2}
\frac{1}{\Delta_1\Delta_2 +K^2} \,,
\label{eq:14}
\end{aligned}
\end{equation}
with
\begin{equation}
  \label{eq:15}
  K = \frac{m k_z}{\sqrt{\epsilon\mu} \kappa  R^2} \left( \frac{1}{p_1^2} - \frac{1}{p_0^2}\right)\,,
\end{equation}
and
\begin{equation}
\begin{aligned}
  \label{eq:16}
  \Delta_1 &=  \frac{I'_m(p_1R)}{p_1 R I_m(p_1R)} -\frac{1}{\epsilon} \frac{K'_m(p_0R)}{p_0R K_m(p_0R)}\,,\\
 \Delta_2 &=  \frac{I'_m(p_1R)}{p_1 R I_m(p_1R)} -\frac{1}{\mu} \frac{K'_m(p_0R)}{p_0R K_m(p_0R)}\,,\\
 \Delta_3 &=  \frac{I'_m(p_1R)}{p_1 R I_m(p_1R)} -\frac{1}{\epsilon} \frac{I'_m(p_0R)}{p_0R I_m(p_0R)}\,,\\
 \Delta_4 &=  \frac{I'_m(p_1R)}{p_1 R I_m(p_1R)} -\frac{1}{\mu} \frac{I'_m(p_0R)}{p_0R I_m(p_0R)} \, ,
\end{aligned}
\end{equation}
where $I_m$ and $K_m$ are Bessel functions, and  $I'_m$ and $K'_m$ their derivatives.
We note that the polarization is not conserved under scattering, i.e.,
$\mathbb{T}^{EH}$, $\mathbb{T}^{HE} \neq 0$.  The scattering Green tensor $\mathbb{\Gamma}$ of the cylinder can be expressed in terms of these matrix elements, following the conventional scattering method \cite{rahi2009}. The comparison to the MSE can be performed by suitable projection. For instance, from the projection $\hat{\bf r}\mathbb{\Gamma}^{EE}\hat{\bf r}'$ on the radial directions $\hat{\bf r}, \hat{\bf r}'$ of $\mathbb{\Gamma}^{EE}({\bf r},{\bf r}')$, all four elements  $\mathbb{T}^{EE}$, $\mathbb{T}^{HH}$, $\mathbb{T}^{HE}$, $\mathbb{T}^{EH}$ can be extracted as they are multiplied by different combinations of $K_m(p_0 r)$, $K'_m(p_0 r)$, $K_m(p_0 r')$, $K'_m(p_0 r')$. Therefore, all components of the analytically computed MSE for $\mathbb{\Gamma}^{EE}$ can be compared to the above T-matrix elements.

\section{Free Green tensors}
\label{sec:green}

For completeness, we provide the explicit expressions of the Green tensors, for a homogeneous and isotropic magneto-dielectric medium with frequency dependent electric and magnetic permittivities $\epsilon_\sigma(\omega)$, $\mu_\sigma(\omega)$, respectively.  
The external sources $({\bf J}, {\bf M})$ are normalized such that  Maxwell Equations for imaginary frequencies $\omega=i \xi$ take the form
\begin{eqnarray}
-{\bf \nabla} \times {\bf E}&=& \kappa \,\mu \;{\bf H} +  {\bf M}\;,\\
{\bf \nabla} \times {\bf H} &=& \kappa\, \epsilon \; {\bf E}+ {\bf J}\;,
\end{eqnarray}
where $\kappa$ is wave number $\kappa=\xi/c$. The components of $6\times 6$ dimensional Green tensor then are
\begin{equation}
\begin{aligned}
\mathbb{G}^{(EE)}_{\sigma,ij}({\bf r},{\bf r}')&=-\frac{1}{\kappa}\left(\frac{1}{\epsilon} \frac{\partial^2}{\partial x_i \partial x'_j} + \mu\, \kappa^2\, \delta_{ij} \right)\, g_\sigma( {\bf r}-{\bf r}')\;, \\
\mathbb{G}^{(HH)}_{\sigma,ij}({\bf r},{\bf r}' )&=-\frac{1}{\kappa}\left(\frac{1}{\mu} \frac{\partial^2}{\partial x_i \partial x'_j} + \epsilon\, \kappa^2\, \delta_{ij} \right)\,  g_\sigma( {\bf r}-{\bf r}')\;, \\
\mathbb{G}^{(HE)}_{\sigma,ij}({\bf r},{\bf r}')&=-\,\epsilon_{ijk} \frac{\partial}{\partial x_k}\,   g_\sigma( {\bf r}-{\bf r}') \;, \\
\mathbb{G}^{(EH)}_{\sigma,ij}({\bf r},{\bf r}')&= -\,\epsilon_{ijk} \frac{\partial}{\partial x'_k}\,  g_\sigma( {\bf r}-{\bf r}')\;,
\label{eq:5}
\end{aligned}
\end{equation}
where  $i,j\in \{x,y,z\}$ denote the spatial components, $\epsilon_{ijk}$ is the Levi-Civita symbol, and the scalar Green function is
\begin{equation}
\label{eq:6}
g_\sigma( {\bf r}-{\bf r}')= \frac{e^{-\kappa \sqrt{\epsilon_\sigma \mu_\sigma} \vert {\bf r}-{\bf r}' \vert }}{4 \pi\,\vert {\bf r}-{\bf r}' \vert}\;.
\end{equation}




\begin{acknowledgments}
Early discussions with B.~Duplantier are acknowledged.
\end{acknowledgments}

\bibliographystyle{apsrev4}



\end{document}